\documentclass[12pt]{article}
\usepackage{natbib}
\usepackage[textwidth=7in,textheight=8.6in,headsep=80pt,footskip=62pt]{geometry}%

\usepackage{amsmath}
\usepackage{amssymb}

\usepackage[font=footnotesize]{caption}

\usepackage{rotating}
\usepackage{url}
\usepackage{graphicx}
\usepackage{dcolumn}
\usepackage{bm}
\usepackage{xcolor} 
\usepackage{bbm}

\newcommand{\new}[1]{{\color{black}  {#1}}}

\newcommand{\vect}[1]{\mbox{\boldmath $#1$}}

\newcommand{\ER}{Erd\H{o}s-R\'{e}nyi }

\newcommand{\LPI}{{\rm LPI}}
\newcommand{\RL}{{\rm RL}}

\begin{document}



\title{Identifying relationship lending in the interbank market:\\ A network approach}

\author{Teruyoshi Kobayashi$^{1}$ and Taro Takaguchi$^{2,3,4}$}

\date{\normalsize{$^1$\emph{Department of Economics, Kobe University, Kobe, Japan} \\
$^2$\emph{National Institute of Information and Communications Technology, Tokyo, Japan} \\
$^3$\emph{National Institute of Informatics, Tokyo, Japan} \\
$^4$\emph{JST, ERATO, Kawarabayashi Large Graph Project, Tokyo, Japan}} \\ [2ex]
\today}
\maketitle

\abstract{
Relationship lending is broadly interpreted as a strong partnership between a lender and a borrower. Nevertheless, we still lack consensus regarding how to quantify the strength of a lending relationship, while simple statistics such as the frequency and volume of loans have been used as proxies in previous studies. Here, we propose statistical tests to identify relationship lending as a significant tie between banks. Application of the proposed method to the Italian interbank networks reveals that the fraction of relationship lending among all bilateral trades has been quite stable and that the relationship lenders tend to impose high interest rates at the time of financial distress.\\
{\bf Keywords}: Relationship lending, interbank markets, temporal networks\\
{\bf JEL Classifications}: G10, G21}

\section{Introduction}

The role of a strong relationship between a lender and a borrower, the so-called relationship lending (or relationship banking), is one of the most widely discussed issues in theoretical and empirical studies of banking. Many empirical studies investigate the economic impact of relationship lending on the terms of loans, such as interest rates and the amount of funds lent, aiming to test the theoretical implications that have been provided since the early 1990s~\citep{Sharpe1990JF,Rajan1992JF,Elyasiani2004JEB,Freixas2008book}. In particular, relationship lending is considered to play an important role in providing liquidity to borrowers facing credit constraints by reducing the extent of information asymmetry between lenders and borrowers. On the other hand, borrowers in relationship trades could be ``locked-in'' by lenders due to their exclusive acquisition of private information, leading to a hold-up problem~\citep{Petersen1995QJE,Von2004Letters,Freixas2008book}.
 
 
There are also a number of studies on relationship lending in interbank markets, where banks lend to and borrow from each other.
The results of the previous analyses, however, are based on ad-hoc and simple measures of relationship lending, and their simplicity may cause a mismeasurement error especially when there is heterogeneity in banks' activities.
A naive measure of relationship lending is the number of transactions between two banks conducted during a certain period of time~\citep{Furfine1999microstructure,Brauning2017RF}. Another widely used measure is the degree of concentration in lending~\citep{Cocco2009JFI,Afonso2013FRB}, measured by the share of funds lent to a particular counterparty. These two measures are expected to straightforwardly capture the strength of a bilateral relationship in the interbank market; a bank pair engaging in relationship lending would trade more frequently and devote a larger share of their total trading volume to the trades between them than to trades with other banks. However, these measures might misinterpret the strength of lending relationships. First, the number of trades with a particular counterparty may merely reflect a bank's need to trade in the interbank market. For instance, if two banks have strong needs to provide and obtain overnight liquidity in the interbank market, respectively, these banks are likely to trade by chance even if they have no preferences for trading partners.  
Second, the degree of concentration in lending can be affected by the difference in the balance-sheet size of counterparties. For example, suppose that a large bank demands a greater amount of funds than smaller banks do. If a bank lends to the large bank, the degree of concentration in lending may appear to be large, even though the lending bank has no preference for partners. The share of lending volume to a particular partner could correctly capture relationship lending if all the counterparties had the same liquidity demands.  Given these limitations, we need a more carefully designed measure of relationship lending that will allow us to control for these factors.
  
In this work, we propose the concept of a \emph{significant tie} as a statistically founded definition of relationship lending. Two banks are said to be connected by a significant tie if the number of trades between them is too large to be explained by random chance after controlling for their intrinsic activity levels.
We control for the activity of banks by employing a simple network-generative model as the null model.
The so-called \emph{fitness model}, one of the standard network-generative models in network science~\citep{Caldarelli2002PRL,DeMasi2006PRE,Musmeci2013JSP}, considers a situation in which the probability of two banks being matched is given as a function of their activity parameters (i.e., \emph{fitnesses}) independently of the history of their transactions. This history-independent property enables us to explicitly compute the theoretical distribution of the number of bilateral trades under the null hypothesis that there is no preference for partners, thus allowing for statistical tests. In this paper, we regard a bank pair connected by a significant tie as engaging in relationship lending.
This definition would eliminate the possible mismeasurement of relationship lending due to differences in banks' activity levels, which should be reflecting their liquidity demands and balance-sheet sizes.

We apply the proposed identification framework to the data on over one million interbank transactions conducted in the Italian interbank market (e-MID) during 2000--2015.
The results reveal important facts about relationship lending, some of which can be summarized as follows. \new{First, throughout the data period, the percentage of relationship lending among all bilateral transactions has been stable, although the percentage slightly increased around the occurrence of particular economic events (e.g., circulation of Euro started in 2002 and the Lehman collapse in 2008).}
  Second, significant ties tend to last for longer periods than non-significant ties do, which is consistent with the conventional notion of relationship lending. Interestingly, the duration of relationships has a decreasing hazard rate (i.e., the probability of ending a relationship is decreasing in duration).
   This implies that the value of relationships in the interbank market increases in time, contrary to the finding of \cite{Ongena2001JFE} on bank--firm relationships.
   Third, the interest rates for relationship lending are indistinguishable from those for transactional lending before and after the global financial crisis, but in the midst of the crisis the borrowers of relationship lending paid significantly higher interest rates than the average. This suggests that some banks faced a hold-up problem at the time of financial distress.
   Fourth, the chance that a bank pair is connected by a significant tie is affected little by the nationality of the banks, suggesting the absence of home-country bias in building bilateral relationships. 
  
\new{The rest of the paper is organized as follows. In section 2, we briefly review related studies and explain about the data on overnight interbank transactions. The method for identifying relationship lending is described in section 3, and the results are shown in section 4. Section 5 provides a robustness analysis and some extensions, and section 6 concludes.}
     
  \new{
 \section{Preliminary}
\subsection{Literature review}  \label{sec:review} 

A large fraction of previous works on relationship lending study bilateral relationships between a bank and a non-financial firm~\citep{Sette2015JEEA,Kysucky2015MS}, while other studies explored the role of relationship lending in the interbank market~\citep{Furfine1999microstructure,Cocco2009JFI,Affinito2012JBF,Craig2015JBF,Hatzopoulos2015,Brauning2017RF}.
For example, \cite{Cocco2009JFI} showed that in the Portuguese interbank market, bilateral trades made by banks with stronger relationships tend to exhibit lower interest rates. In Italy, \cite{Affinito2012JBF} found that relationship lenders played an essential role as liquidity providers, especially in the midst of the global financial crisis of 2007--2009. \cite{Brauning2017RF} argued that during the financial crisis, relationship lenders in Germany offered lower interest rates to their close partners.  \cite{Hatzopoulos2015} proposed a null model based on a hypergeometric distribution for testing the significance of edges in the e-MID market.

 In the literature, measuring the influence of Lehman bankruptcy on the interbank market has been one of the central interests.
  \cite{afonso2011stressed} argue that counter party risk became more important than liquidity hoarding at the time of Lehman Brothers' collapse, showing that loan terms got more sensitive to borrowers' credit worthiness. \cite{angelini2011interbank} also show that the risk of moral hazard due to ``too-big-to-fail'' increased during the crisis compared to the period prior to August 2007.

 The e-MID market has also been extensively studied from a point of view of complex networks. \cite{Iori2008JEDC}, \cite{finger2013network} and \cite{Fricke2015CompEcon} analyzed the topology of aggregated interbank networks, while  \cite{Barucca2016soliton} and focused on the time-varying nature of interbank networks. \cite{Kobayashi2017arxiv} find several temporal patterns in bilateral transactions that are similar to the ones observed in social communication patterns of humans~\citep{Cattuto2010PLOS}. Examples of studies of other interbank markets include \cite{craig2014interbank} for Germany, \cite{giraitis2016estimating} for the UK, \cite{Cont2013} for Brazil, and \cite{Imakubo2010BOJ} for Japan. 
}

\subsection{Data}

\begin{figure}[t]
    \centering
       \includegraphics[width=.7\columnwidth]{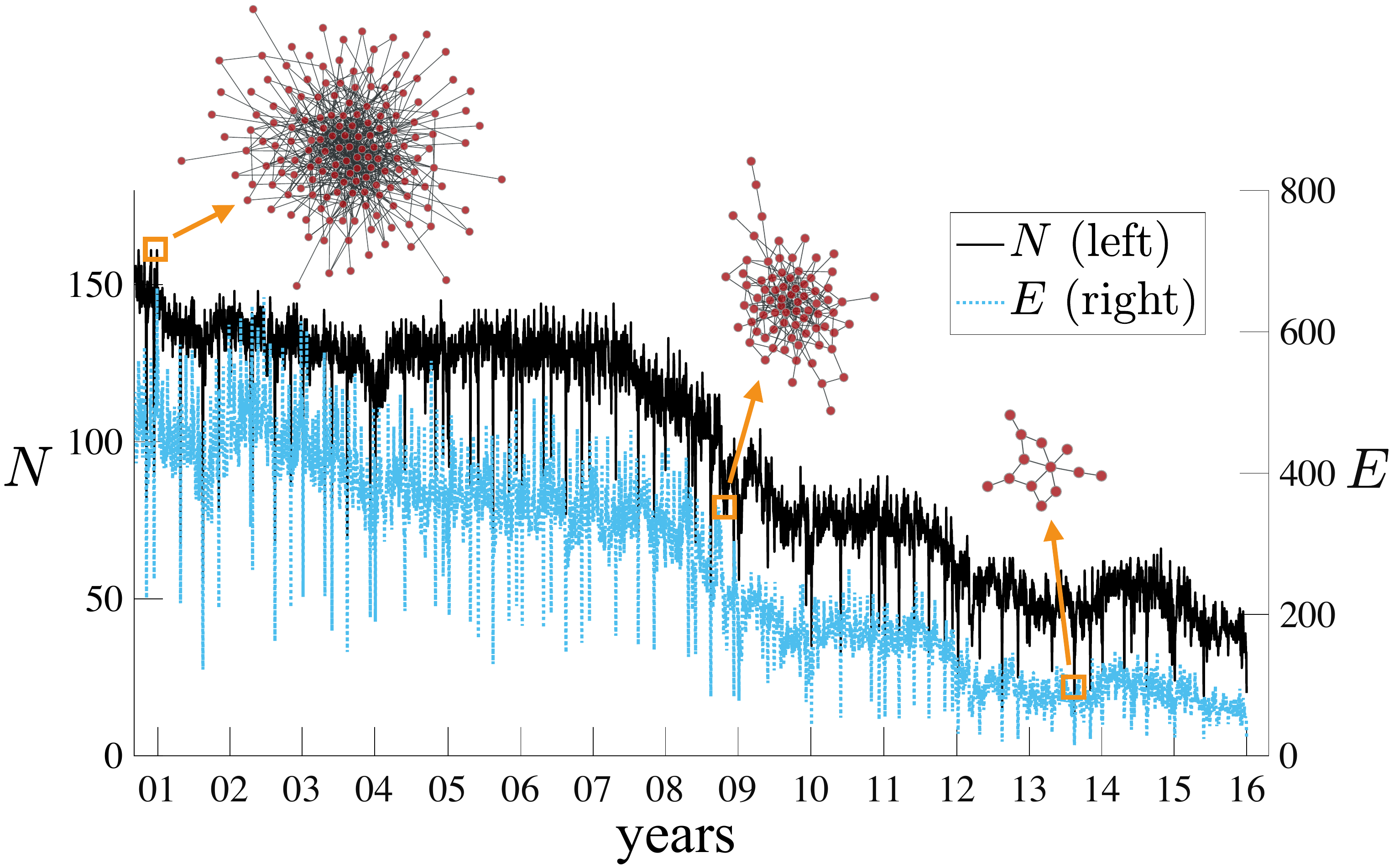}
    \caption{\footnotesize{Evolution of interbank networks. Solid and dotted denote the number of active banks $N$ (left axis) and the number of edges $E$ (right axis) of a daily network, respectively. The largest (November 23, 2000), a middle-sized (October 29, 2008) and the smallest networks (August 15, 2013) are visualized.} }
    \label{fig:ENplot}
\end{figure}

We use time-stamped data on interbank transactions conducted in the Italian online interbank market (e-MID) between September 2000 and December 2015.
\new{As in the other interbank markets, e-MID plays a role as a marketplace in which banks in need of short-term liquidity or having excess liquidities find counterparties by posting an order on the platform. Banks that post requests are called \emph{quoters}, and their counterparties are called \emph{aggressors}. The actual names of trading banks are not revealed in the platform, but their proper IDs, including their nationality, are made public (e.g., ``\url{IT0002}'', where ``\url{IT}'' denotes Italy).
The transactions data contain the following information: date and time (e.g., ``\url{2000-09-04}\ \url{09:12:40}''), the IDs of banks, maturity,  interest rates, and trade amount (in million Euros). The e-MID data is commercially available from e-MID SIM S.p.A based in Milan, Italy (\url{http://www.e-mid.it/}). }

  In this work, we use the overnight transactions of unsecured Euro deposits labeled as ``ON" (i.e., overnight) or ``ONL" (i.e., overnight large, namely overnight transactions no less than 100 million Euros), which comprise the great majority of transactions ($> 86\%$) in the e-MID market.  
   An advantage of focusing on overnight trades is that we can construct a sequence of snapshots of daily interbank networks having banks as nodes and lending-borrowing relationships as edges (Fig.~\ref{fig:ENplot}). An edge is created when a loan is executed. If there are multiple transactions between two banks during a day, we represent the trading relationship as one unweighted edge. As a result, the number of edges over the whole data period totals 1,033,349. 
  
From Fig.~\ref{fig:ENplot}, it is evident that interbank networks constantly change their size on a daily basis, and there is a common downward trend in the numbers of active banks $N$ and edges $E$. Here, ``active'' banks in a daily network are defined as banks that had transactions at least once between 9:00 and 18:00. 
  Downward spikes in $N$ and $E$ are mostly due to national holidays in Italy.\footnote{Weekends are not included since the market is closed.} On the other hand, the presence of a long-term downward trend could be attributed to multiple factors such as the onset of the global financial crisis, the Greek sovereign debt crisis, and the introduction of highly expansionary monetary policies of the the European Central Bank (ECB) (and possibly other central banks). Summary statistics of the time series of daily interbank networks are presented in Table~\ref{tab:summary_stats}. 
 
\begin{table}[]
    \caption{Summary statistics of the daily interbank networks. Symbol $\overline{x}$ denotes the average of variable $x$ over the corresponding period, and $\langle k \rangle$ is the daily average degree. Subscripts ``max'' and ``min'' represent the maximum and minimum values, respectively.}
    \begin{center}
    \begin{tabular}{lcccc}
    \hline
    \hline
       & All & \hspace{.1cm}2000--2006 & \hspace{.1cm}2007--2009 & \hspace{.1cm}2010--2015 \\ 
       \hline
        \# days &3,922  &1,618  &767& 1,537  \\
       $\overline{N}$  &95.80 &130.40  & 101.67& 56.45   \\
         $N_{\rm max}$ &161 &161 &144 & 89  \\
        $N_{\rm min}$ &13  &56 &48 & 13  \\
         $\overline{E}$ &262.96  &402.16 &266.00 & 114.91  \\
        $E_{\rm max}$ &662  &662 &461 & 265  \\
        $E_{\rm min}$ &15  &122 &76 &  15\\
        $\overline{\langle k\rangle}$ &2.54  & 3.07&  2.57& 1.97 \\
        \hline
    \end{tabular}\\
     \label{tab:summary_stats}
         \end{center}
\end{table}

\section{Model and methods}
\label{sec:model_methods}

\subsection{Fitness model}

As a baseline framework for the subsequent statistical analysis, we introduce here a simple model of daily interbank networks that describes how a lender and a borrower are matched. Our model is a variant of the fitness model~\citep{Caldarelli2002PRL,DeMasi2006PRE}. The fitness model has been frequently used in the field of network science to explain the mechanism of dynamic network formation, in which the probability that two agents are connected depends on the fitness of the agents.  In the context of interbank markets, fitness corresponds to the intrinsic activity level of a bank, such as the demand for short-term liquidity if the bank is a possible borrower and the willingness to supply funds if the bank is a possible lender. In spite of its simplicity, the fitness model has been shown to explain many rich properties that emerge from the evolution of interbank networks~\citep{DeMasi2006PRE,Kobayashi2017arxiv}.

In the baseline model, we regard daily interbank networks as undirected (i.e., we ignore the direction of edges) because our main focus is on identifying and analyzing the role of the bilateral relationship between banks. \new{We will extend the analysis to directed networks in section~\ref{sec:directed}.} 
We assume that the probability $u$ that bank $i$ trades with bank $j$ on a given day is expressed by the product of their activity levels:
\begin{align}
u(a_{i},a_{j}) \equiv a_{i} a_{j},    \label{eq:matching_func}
\end{align} 
where $a_{i} > 0$ represents the activity level (or fitness) of bank $i$.\footnote{In \cite{Kobayashi2017arxiv}, we used a matching function of the form $u(a_{i}, a_{j}) = (a_{i}a_{j})^{\alpha}$. In the current model, we can set $\alpha = 1$ without loss of generality because the case of $\alpha \neq 1$ can be recovered by redefining the activity parameter as $a^{\alpha}$.} 
The model nests a wide variety of well-known network generating models, depending on the specification of $\{a_{i}\}$.
For example, if $a_i = a \; \forall \:i$, then the model is equivalent to an \ER random graph with constant matching probability $u = a^{2}$~\citep{Erdos1959PublMath}. If $a_i = k_{i}/\sqrt{2M}$, where $k_i$ and $M$ are the degree of bank $i$ and the total number of edges in a daily network, respectively, then the matching probability is given by $u = k_{i}k_{j}/(2M)$, resulting in the configuration model~\citep{Newman2010book}.\footnote{The configuration model is a network model that generates a random network having a predefined degree sequence $\{k_i\}$. See \cite{Newman2010book} for details.} 

 We first estimate the activity vector $\vect{a}\equiv (a_1,\ldots , a_N)$, assuming that every element of $\vect{a}$ is constant during an aggregate period consisting of $\tau$ consecutive business days. In other words, daily networks in an aggregate period are regarded as independent realizations from the fitness model with estimated $\vect{a}$. \new{In section~\ref{sec:time-varying}, we will consider the case of time-varying activity parameters.} 
In short, we are extracting a $N\times 1$ vector of bank activity levels from the observed network structure containing $N\times (N-1)$ elements of information on bilateral trades (i.e., adjacency matrix).  
This dimensionality reduction obviously discards the structural information of a network. 
In return, the resultant estimates enable us to infer the extent to which a random matching between banks can explain the empirical network structure, avoiding an over-identification problem. Based on the estimates of $\vect{a}$, we identify the existence of relationship lending by testing whether the observed number of transactions between two banks is significantly larger than the value expected by the null hypothesis (i.e., the fitness model).


\subsection{Maximum likelihood estimation of activity levels}
\label{sec:MLE}

We split the daily data set into aggregate periods, each consisting of $\tau$ business days, and perform a maximum likelihood estimation of $\vect{a}$ period by period.
Aggregate periods are indexed by $t^\prime = 1,\ldots, t_{\rm max}^\prime$, where $t_{\rm max}^\prime \equiv \lfloor t_{\rm max} / \tau \rfloor$ and $t_{\rm max}$ denotes the total number of business days in the data. For the sake of simplicity, we omit subscript $t^\prime$ in the rest of this section.

 If trading pairs are independently matched each day according to probability $u(a, a^\prime)$, then the number of trades between banks $i$ and $j$ conducted over $\tau$ business days follows a binomial distribution with parameters $\tau$ and $u(a_{i},a_{j})$. For a given activity vector $\vect{a}$, the joint probability function of the number of trades in an aggregate period then leads to
\begin{align}
   p(\{m_{ij}\}|\vect{a}) = \prod_{i,j: i\neq j}\begin{pmatrix} \tau \\ m_{ij}\end{pmatrix} u(a_{i},a_{j})^{m_{ij}} (1-u(a_{i},a_{j}))^{\tau-m_{ij}},
\end{align} 
where $m_{ij}\leq \tau$ denotes the number of trades (i.e., edges) between $i$ and $j$ observed in an aggregate period. The log-likelihood function is thus given by
\begin{align}
   \mathcal{L}(\vect{a}) & = \log p(\{m_{ij}\}|\vect{a}) \notag \\
    &= \sum_{i,j: i\neq j}\left[  m_{ij}\log{(a_{i}a_{j})} + (\tau-m_{ij}) \log{(1-(a_{i}a_{j}))}\right] + \text{const.},
   \label{eq:loglikelihood}
\end{align} 
where ``$\text{const.}$'' denotes the terms that are independent of $\vect{a}$. Let $N$ denote the number of active banks that have at least one transaction during a given aggregate period. 
The maximum-likelihood estimate of $\vect{a}$ is the solution for the following $N$ equations:
\begin{align}
 H_i(\vect{a}^{*}) \equiv & \sum_{j:j\neq i}\frac{m_{ij}-\tau(a_{i}^{*}a_{j}^{*})}{1-(a_{i}^{*}a_{j}^{*})} = 0, \; \forall \: i = 1,\ldots, N, \label{eq:ML_a} 
 \end{align}
The first-order condition \eqref{eq:ML_a} is obtained by differentiating the log-likelihood function Eq.~\eqref{eq:loglikelihood} with respect to $a_i$. The system of nonlinear equations, $H(\vect{a})=\vect{0}$, can be solved by using a standard numerical algorithm.\footnote{We solved the problem by using the Matlab function \url{fsolve}, which is based on a modified Newton method, called the trust-region-dogleg method. The initial values of $\vect{a}$ are given by the configuration model, $a_i = \sum_{j:j\neq i}(m_{ij}/\tau)/\sqrt{2\sum_{i<j}m_{ij}/\tau}$, where the numerator and the denominator represent the daily means of bank $i$'s degree and the doubled number of total edges, respectively. There are a few cases in which the estimated activity values $a_{i}$ and $a_{j}$ indicate $u(a_{i},a_{j})>1$. In such cases, we assume $u$ = 1.} 
Hereafter, the computed solution (i.e., the maximum likelihood estimate) of $\vect{a}$ is denoted by $\vect{a}^*\equiv (a_1^*,\ldots , a_N^*)$.
By repeating this process period by period, we obtain the estimates of activity vectors $\left\{ \vect{a}^*_1, \vect{a}^*_2, \ldots, \vect{a}^*_{t_{\rm max}^\prime} \right\}$.

\subsection{Statistical tests for relationship lending}
\label{sec:stat_tests}

 Here we present two sorts of statistical tests; one is for identifying bank pairs engaging in relationship lending and the other is for detecting relationship-dependent banks.
 In the same manner as we estimate the activity levels of banks (section~\ref{sec:MLE}), we split the daily data set into $t_{\rm max}^{\prime}$ aggregate periods and implement the tests period by period.

\subsubsection{Edge-based test for relationship lending}\label{sec:edge_test}

If bank $i$ has no preference for trading partners and thereby finds a partner in a random manner as suggested in the fitness model, then the number of bilateral transactions between banks $i$ and $j$ during a given period, $m_{ij}$, should obey the following binomial distribution: 
 \begin{align}
   g(m_{ij}|a_{i}^*,a_{j}^*) = \begin{pmatrix} \tau \\ m_{ij}\end{pmatrix} u(a_{i}^*,a_{j}^*)^{m_{ij}} (1-u(a_i^*,a_j^*))^{\tau-m_{ij}},  \ \forall \: i,j = 1,\ldots, N.
   \label{eq:edge_test}
\end{align} 

\begin{figure}[t]
    \centering
       \includegraphics[width=.65\columnwidth]{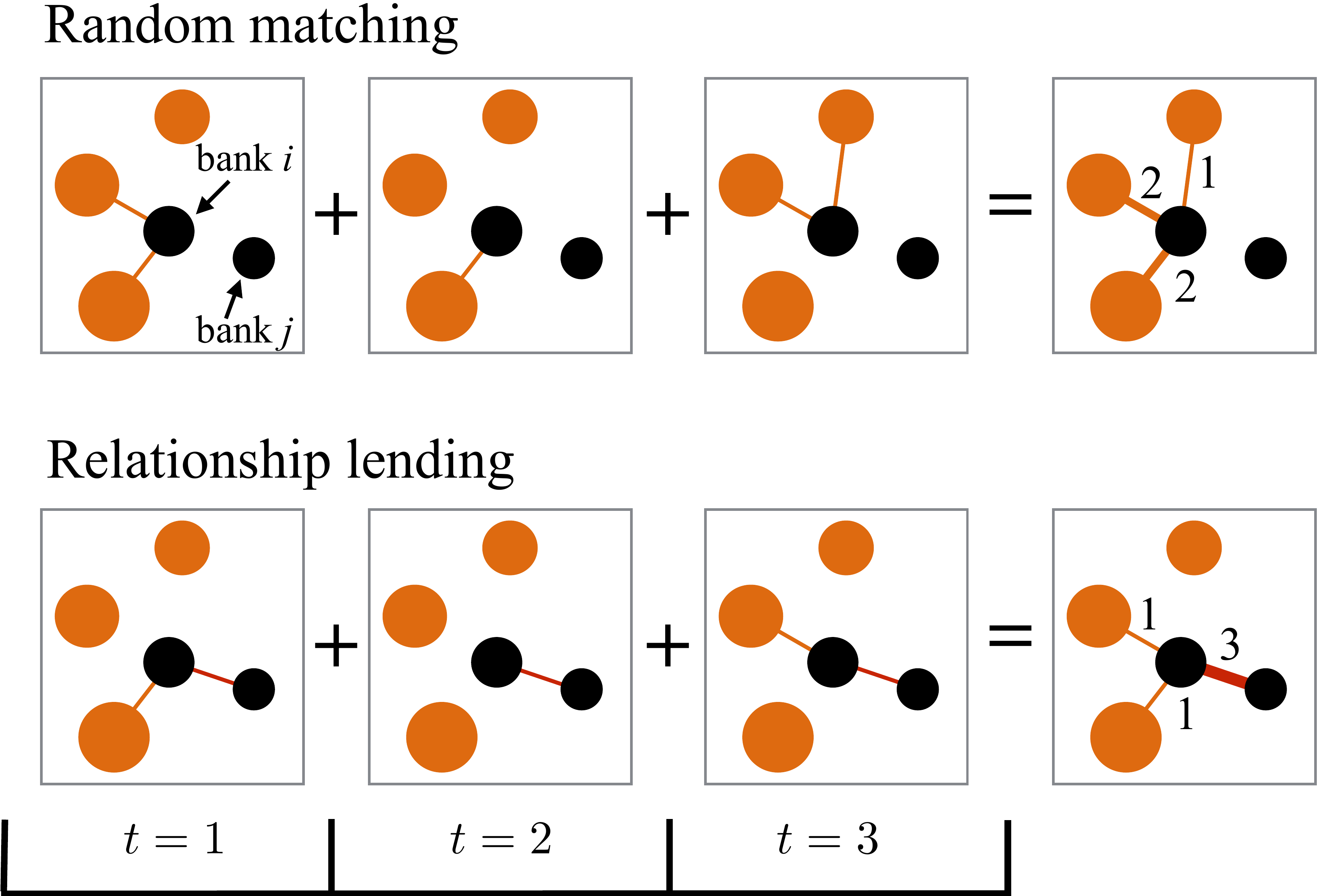}
    \caption{Schematic of a significant tie. For illustrative purposes we set $\tau=3$. The size of circle represents the activity level of a bank. If banks are matched randomly according to the fitness model, then banks with higher activity levels will receive larger number of edges on average. If the number of trades between bank $i$ and bank $j$ is too large to be explained by random chance, then the two banks are considered to be connected by a significant tie and engaging in relationship lending.}
       \label{fig:schematic_sigtie}
\end{figure}

In contrast, if bank $i$ has a strong (i.e., non-random) partnership with bank $j$, then the distribution of $m_{ij}$ will deviate from a binomial distribution. 
Let $m_{ij}^{c}$ denote the $c$-th percentile $(0 \leq c \leq 100)$ of $g(m_{ij}|a_{i}^*,a_{j}^*)$ (i.e., $c/100 = G(m_{ij}^{c} |a_{i}^*,a_{j}^*)$), where $G$ is the cumulative distribution function (CDF) of $g(m_{ij}|a_{i}^*,a_{j}^*)$. If $m_{ij} > m_{ij}^{c}$ for a $c$ value close to $100$, then the empirical number of transactions is too large to be explained by random chance, indicating the presence of relationship lending. We call this test the \emph{edge-based test} since this is a test for the significance of edges in interbank networks. 
If $m_{ij} > m_{ij}^{c}$, then we say that banks $i$ and $j$ are connected by a significant tie and engaging in relationship lending. We set $c=99$ (i.e., 99\% significance level) throughout the paper. A schematic of a significant tie is presented in Fig.~\ref{fig:schematic_sigtie}. 

Importantly, the number of bilateral trades in a given period itself does not necessarily indicate the presence of a significant tie. 
Under a random matching (the upper row of Fig.~\ref{fig:schematic_sigtie}), bank~$i$ trades twice with each of the two counterparties having the largest activity levels, which should be a natural consequence given the high matching probabilities. By contrast, banks $i$ and $j$ trade three times in the bottom row of Fig.~\ref{fig:schematic_sigtie}, which is unexpected based on their small activity levels. Therefore, bank $i$ is considered to engage in relationship lending with bank $j$ but not with the other three.

\subsubsection{Node-based test for relationship-dependent banks}\label{sec:node_test}

Since we have random matching probabilities $u(a, a^\prime)$ for any pairs of banks, we can also test the extent to which a bank depends on a limited number of partners.
The probability function of aggregate degree $K_i$ is given as 
\begin{align}
   f(K_i|\vect{a}^*) &= \sum_{\left\{ A_{ij} \right\}} \prod_{j:j\neq i} g(m_{ij} = 0)^{1- A_{ij}} (1-g(m_{ij} = 0))^{A_{ij}} \times \delta\left( \sum_j A_{ij}, K_i\right) \notag \\
   &=  \sum_{\left\{ A_{ij} \right\}} \prod_{j:j\neq i}(1-u(a_i^*,a_j^*))^{\tau(1-A_{ij})} (1-(1-u(a_{i}^*,a_{j}^*))^{\tau})^{A_{ij}} \times \delta\left( \sum_{j} A_{ij}, K_{i}\right),
\label{eq:df_deg}
\end{align}
where $A_{ij}$ is the $(i,j)$-element of the aggregate adjacency matrix; $A_{ij} = 1$ if there is at least one transaction between banks $i$ and $j$ during an aggregate period, and $A_{ij}= 0$ otherwise.  $\delta(x,y)$ denotes the Kronecker delta which equals one if $x = y$ and zero otherwise. Note that the second equality follows from relation $g(m_{ij} = 0) = (1-u(a_i^*,a_j^*))^{\tau}$ (Eq.~\eqref{eq:edge_test}).
In fact, Eq.~\eqref{eq:df_deg} is equivalent to the distribution of the sum of $N-1$ random variables drawn from a Bernoulli distribution with parameter $\{1-(1-u(a_{i}^*,a_{j}^*))^{\tau}\}_{j:j\neq i}$, or a Poisson binomial distribution. 
Here we would like to compute the CDF of $f(K_i|\vect{a}^*)$ to evaluate the significance of empirical $K_i$. However, exact calculation of the CDF of a Poisson binomial distribution is notoriously difficult because one must compute $\binom{N}{K_i}$ number of terms~\citep{Steele1994LeCam}.
Thus, we instead approximate the probability distribution of $K_i$ to a Poisson distribution~\citep{LeCam1960}:
\begin{align}
   f(K_i|\vect{a}^*) \approx \frac{\lambda_i^{* K_i}e^{-{\lambda_i^{*}}}}{K_{i}!} \equiv \widetilde{f}(K_i|\vect{a}^*), 
\label{eq:df_poiss}
\end{align} 
where $\lambda_{i}^{*} \equiv \sum_{j:j\neq i}[1-(1-u(a_{i}^*,a_{j}^*))^{\tau}]$. An error bound for this Poisson approximation is provided by an extended version of the Le Cam's theorem~\citep{LeCam1960,Barbour1983poisson,Steele1994LeCam}:
\begin{align}
\sum_{K_i=0}^\infty\left| f(K_i|\vect{a}^*) - \frac{\lambda_{i}^{*K_i}e^{-\lambda_i^{*}}}{K_{i}!}\right| < \frac{2(1-e^{-\lambda_{i}^{*}})}{\lambda_{i}^{*}}\sum_{j:j\neq i}p_{ij}^2,  \; \forall \; i,j,
\label{eq:LeCam}
\end{align}
where $p_{ij} \equiv 1-(1-u(a_{i}^*,a_{j}^*))^\tau$. 

The Poisson approximation enables us to formally test the null hypothesis that the empirical aggregate degree $K_i$ is explained by random chance.
Let $K_i^{c^\prime}$ denote the $c^\prime$-th percentile $(0 \leq c^\prime \leq 100)$ of $\widetilde{f}(K_i|\vect{a}^*)$.
In other words, $c^\prime /100 = \widetilde{F}(K_i|\vect{a}^*)$, where $\widetilde{F}(K_i|\vect{a}^*)$ is the CDF of $\widetilde{f}(K_i|\vect{a}^*)$.
If the data reveal that $K_i < K_i^{c^\prime}$ for $c^{\prime}$ close to zero, then bank $i$ has a significantly smaller number of trading partners than random chance would suggest. If this is the case, it indicates a significant dependence of bank $i$ on relationship lending. Hereafter we call this type of test the \emph{node-based test}, and we set $c^\prime = 1$.

\subsection{Selection of aggregate length $\tau$}
\label{sec:selection_tau}

Before applying the model and statistical tests described in the previous sections to empirical data, we must determine parameter $\tau$, the length of an aggregate period.
In fact, varying $\tau$ would cause trade-offs between approximation accuracy and the stability of aggregated data. 
On the one hand, the choice of $\tau$ would directly affect the accuracy of the Poisson approximation through its influence on $\lambda_i^*$ and $p_{ij}$ in Eq.~\eqref{eq:LeCam}. The average error bound (Eq.~\eqref{eq:LeCam}) increases with $\tau$ as  $\lim_{\tau\to\infty}{p_{ij}}=1, \forall\; i,j$ (Fig.~\ref{fig:select_tau}a in Supplementary Information). Taking into account this positive relationship between the error bound and $\tau$, $\tau$ should be set as small as possible.   
On the other hand, employing a smaller value of $\tau$ would also affect the stability of statistical results as the aggregate networks could become more unstable because the number of active banks would change drastically period to period (see Fig.~\ref{fig:ENplot}). This necessarily reduces the stability of the data to be examined.
Figure~\ref{fig:select_tau}b illustrates that the average and the standard deviation of the absolute changes in $N$, denoted by $\Delta N_{t^{\prime}} \equiv |N_{t^{\prime}} - N_{t^{\prime}-1}|$ $(t^\prime = 2, \ldots, t_{\rm max}^\prime)$, take minimum values around $\tau = 12$. 
Judging from these observations, we employ $\tau = 10$ as a benchmark value. We will show that all the qualitative results shown in this paper are quite robust and not sensitive to the choice of $\tau$.

\section{Results}
  
\subsection{Estimation results: Activity level}

 \begin{figure}[t]
    \centering
       \includegraphics[width=.6\columnwidth]{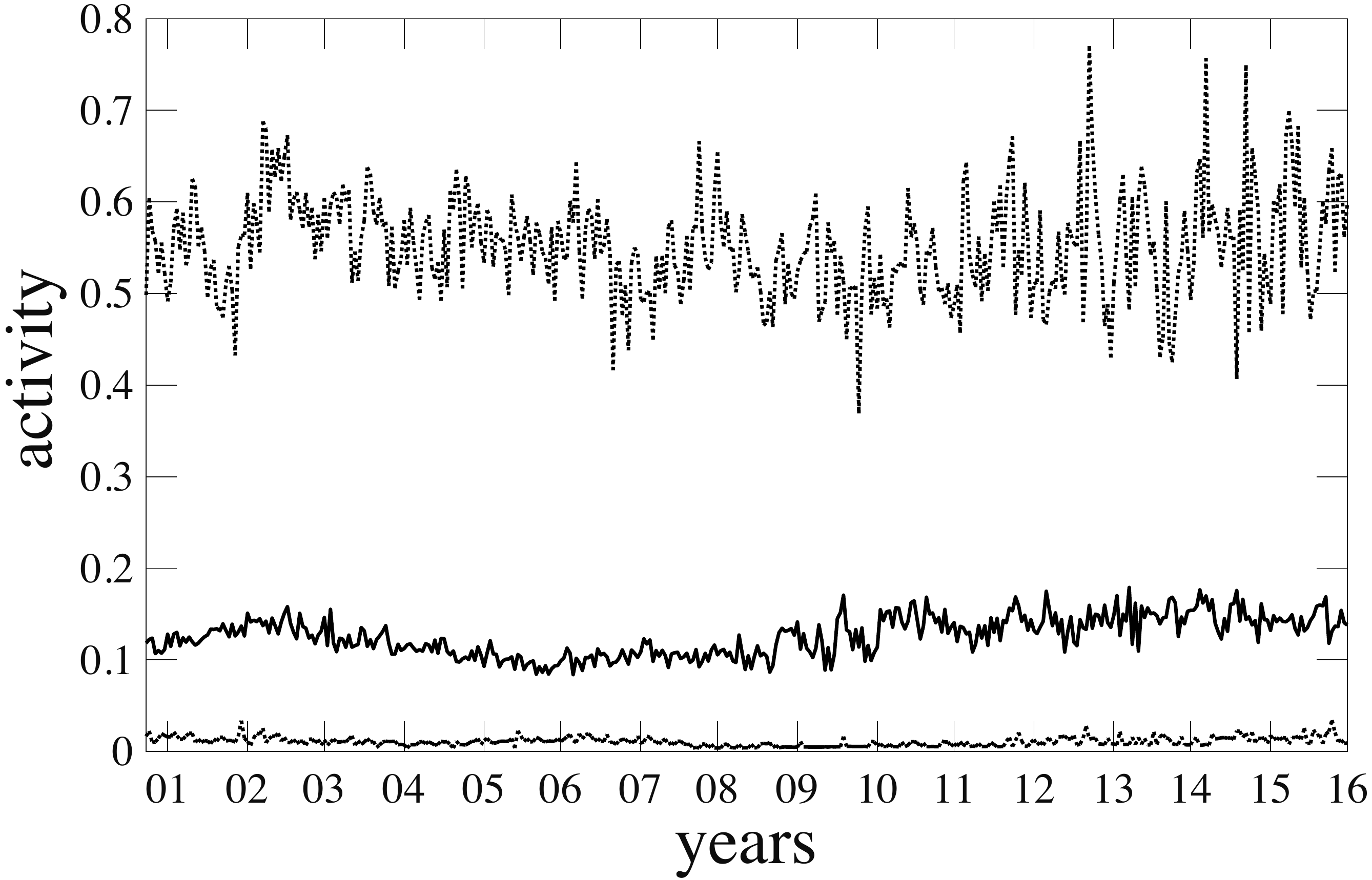}
    \caption{Maximum-likelihood estimates of activity. Solid line represents median, and lower and upper dotted lines respectively denote the 5th and 95th percentiles of the estimated activity distribution in each aggregate period.}
    \label{fig:ML_params}
\end{figure}

\begin{figure}[t]
    \centering
       \includegraphics[width=.8\columnwidth]{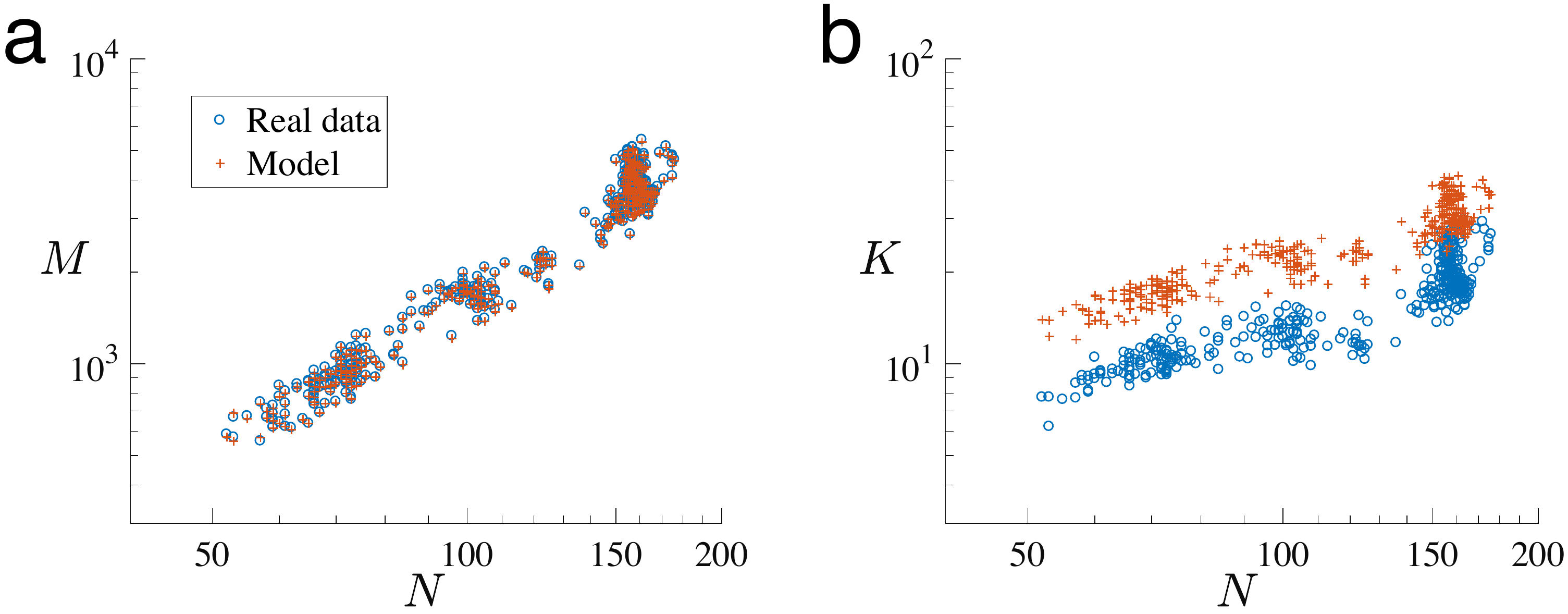}
           \caption{Comparison between the model and the empirical values of $M$ and $K$. Each dot corresponds to an aggregate period. (a) The total number of edges predicted by the model (Eq.~\eqref{eq:Mstar}) well fits the empirical data. (b) The number of unique partners predicted by the model (Eq.~\eqref{eq:Kstar}) overestimates the empirical data.}
    \label{fig:NM_scaling}
\end{figure}

 The distribution of the estimated activity levels, $\vect{a}^{*}$, is shown in Fig.~\ref{fig:ML_params}. The distribution has been relatively stable throughout the data period. Based on these estimates, we can infer how many transactions would be conducted under the null hypothesis in which the matching probability is given by $u(a_{i}^{*},a_{j}^{*})\: \forall \: i,j$. The empirical number of transactions in an aggregate period, denoted by $M$, is given as
 \begin{align}
  M = \sum_{i < j} m_{ij}. 
 \end{align}
 The expected number of transactions under the null hypothesis $M^{*}$ is given as
 \begin{align}
  M^{*} = \tau \sum_{i < j} u(a_{i}^{*},a_{j}^{*}).  \label{eq:Mstar}
 \end{align}
  Figure~\ref{fig:NM_scaling}a illustrates the relationships between $N$ and $M$ in the empirical data and the estimated model. The almost perfect fit between the estimated values of $M^{*}$ and the empirical data indicates that the maximum likelihood estimation works fairly well; the estimated activity accurately captures the actual bank activity in terms of the total number of trades.  In \cite{Kobayashi2017arxiv}, we showed that there is a clear superlinear relationship between the numbers of banks and edges at the daily scale (i.e., $\tau=1$) using the same data. Figure~\ref{fig:NM_scaling}a in fact reveals that a similar scaling relation arises even at the aggregate level of $\tau=10$ business days. 

On the other hand, if we take the presence of relationship lending as a given, the empirical numbers of trading partners should be smaller than the estimated values under the null hypothesis.
To see this,  Fig.~\ref{fig:NM_scaling}b shows the average of aggregate degree $K$, the number of unique trading partners in an aggregate period: 
\begin{align}
 K = \frac{1}{N}\sum_{i,j}A_{ij}.
\end{align}
Under the null hypothesis, the average aggregate degree is computed as
\begin{align}
 K^{*} = \frac{1}{N}\sum_{i,j}[1-(1-u(a_{i}^{*},a_{j}^{*}))^{\tau}]. \label{eq:Kstar}
\end{align}
As shown in Fig.~\ref{fig:NM_scaling}b, $K^*$ overestimates $K$, meaning that in the real world banks tend to be more selective than a random matching would suggest.
In the next section, we identify the presence of relationship lending by statistically testing the extent of deviation from the null model.

\subsection{Identification results: Significant ties and relationship-dependent banks}

\begin{figure}[t]
    \centering
     \includegraphics[width=.85\columnwidth]{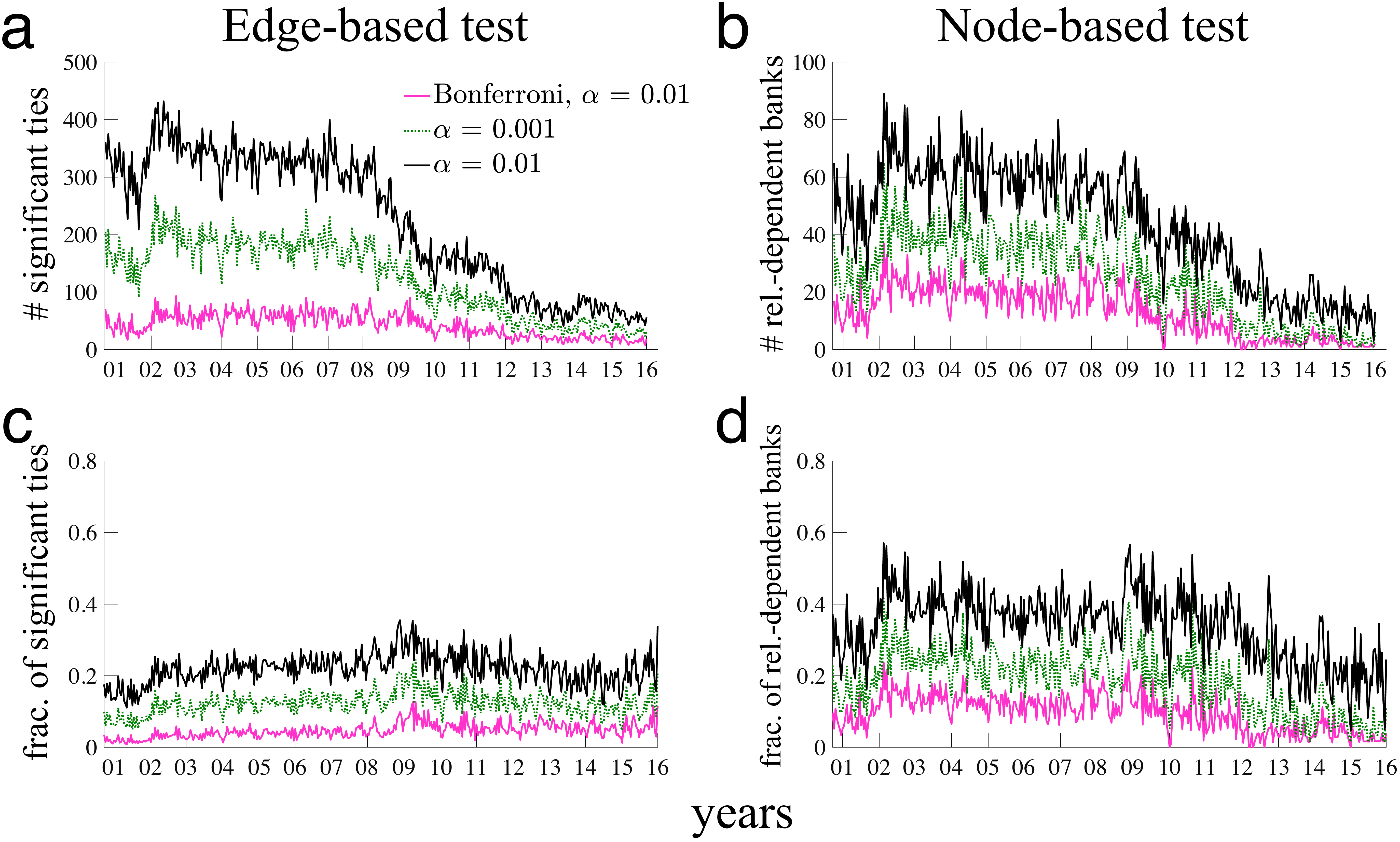}
    \caption{Statistical identification of relationship lending. The numbers of (a) significant ties and (b) relationship-dependent nodes. The fraction of (c) significant ties and (d) relationship-dependent nodes. \new{$\alpha\equiv 1-c/100$ denotes the significance level. ``Bonferroni'' (pale-red line) denotes the Bonferroni correction for multiple statistical tests.}}
    \label{fig:frac_pref} 
    \end{figure}
    
Figures~\ref{fig:frac_pref}a and \ref{fig:frac_pref}c, respectively, show the number and the fraction of significant ties identified by the edge-based test. We also checked the robustness of the results to different choices of $\tau$ in SI (Fig.~\ref{fig:frac_pref_robustcheck}).
Overall, while the number of significant ties has been decreasing along with the downward trend of $E$ (see Fig.~\ref{fig:ENplot}), \new{the percentage of significant ties among all ties is relatively constant for a given level of statistical significance.
 However, we see that the fraction of significant ties apparently went up at the beginning of 2002, when the circulation of Italian lira officially ended, and after the collapse of Lehman Brothers in October 2008.}
 
 
Figures~\ref{fig:frac_pref}b and \ref{fig:frac_pref}d show the number and share of relationship-dependent banks identified by the node-based tests, respectively. As in the case of significant ties, the share of relationship-dependent banks increased drastically at the beginning of 2002 and after the failure of Lehman Brothers. We note that the results of the node-based tests should be treated with care; the fraction of relationship-dependent banks increases with $\tau$ while the fraction of significant ties is almost unaffected (Fig.~\ref{fig:frac_pref_robustcheck}). 
A possible reason for this dependence on $\tau$ is a deterioration in the accuracy of the Poisson approximation (Eq.~\eqref{eq:LeCam}) as described in section~\ref{sec:selection_tau}. Although the absolute values of the fraction of relationship-dependent banks vary with $\tau$, the relative trends over the data period appear still similar.

As we saw in Fig.~\ref{fig:frac_pref}, the proposed methods (section~\ref{sec:stat_tests}) allow us to statistically identify bank pairs engaging in relationship lending and relationship-dependent banks. It is worth noting that this would not be possible without an appropriate null model, which was missing in previous studies (see section \ref{sec:comparison} for an evaluation of the previous measures for the strength of relationship lending).

Information regarding banks' country IDs enables us to investigate the correlations between banks' nationality and the existence of a significant tie and between nationality and the chance of being a relationship-dependent bank. Since Italian banks occupy a great majority in the e-MID market, we split all ties into three combinations of nationalities: Italian-Italian, Italian-foreign, and foreign-foreign pairs.\footnote{The list of all countries is as follows (the number of banks is in parenthesis): Austria (2), Belgium (6), Switzerland (6), Germany (23), Denmark (1), Spain (7), Finland (1), France (10), Great Britain (14), Greece (6), Ireland (5), Italy (213), Luxembourg (4), Holland (4), Norway (1), Poland (1), and Portugal (4).} 
 
   \begin{figure}[t]
    \centering
     \includegraphics[width=.8\columnwidth]{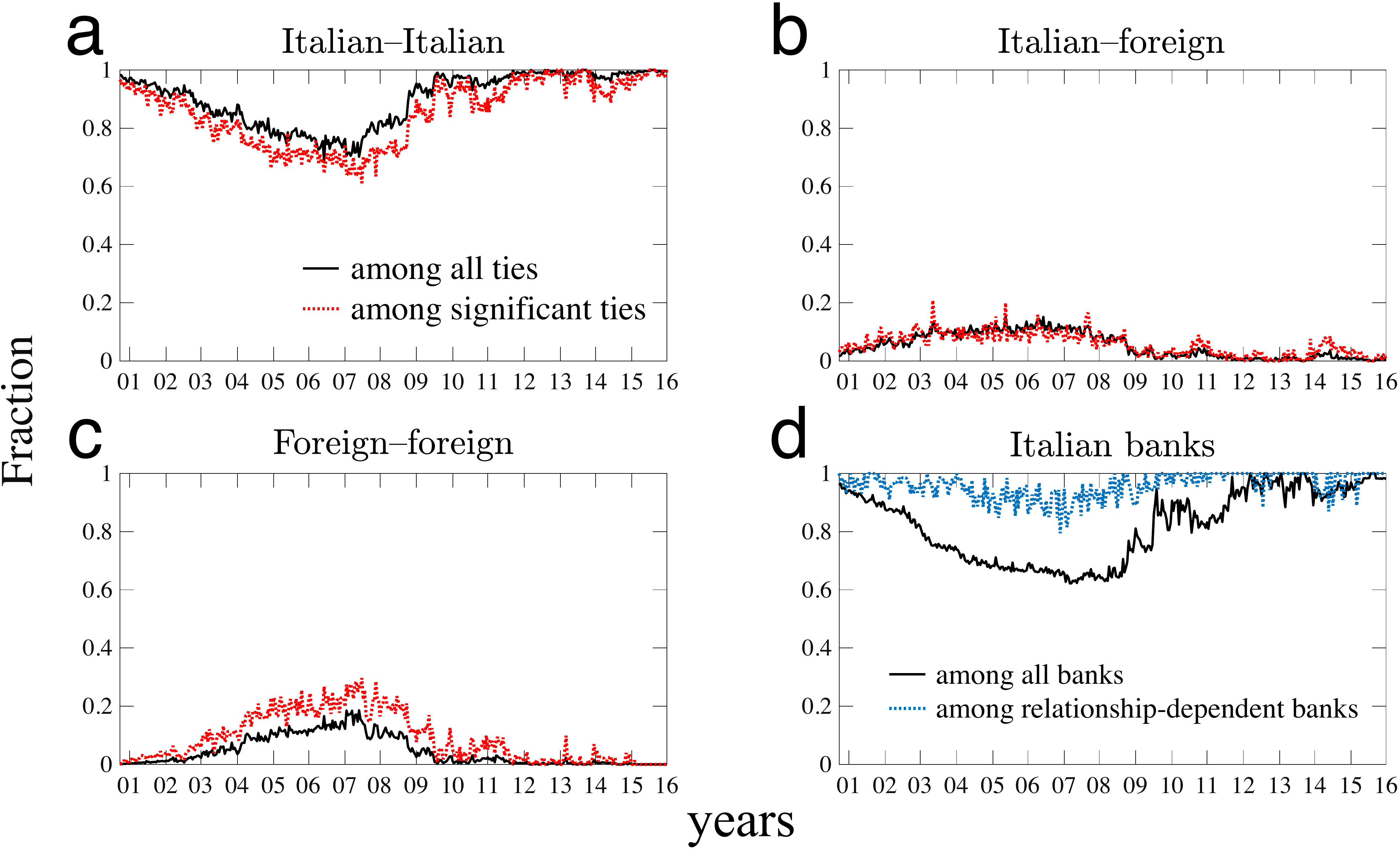}
    \caption{Fraction of (a) Italian-Italian, (b) Italian-foreign, and (c) foreign-foreign bank pairs. Solid and dotted lines indicate the fractions of the corresponding pairs among all pairs and among relationship pairs, respectively. (d) Fraction of Italian banks among all banks (solid) and among relationship-dependent banks (dotted). \new{99\% significance level}.}
    \label{fig:Italian_foreign}
    \end{figure}

 As shown in Fig.~\ref{fig:Italian_foreign}, the fraction of Italian-Italian pairs among all pairs was close to one in the early 2000s, yet it considerably decreased toward the onset of the global financial crisis in 2007--2008.  At the same time, Italian-foreign and foreign-foreign pairs started to increase their presence over the pre-crisis period.  The fraction of Italian-Italian pairs began to increase again shortly after the financial crisis occurred, gradually returning to its pre-crisis level. This seems to suggest that an Italian bank tends to trade with other Italian banks when the market is under stress. However, the share of Italian-Italian significant ties among all significant ties moved in sync with the fraction of Italian-Italian pairs among all pairs, suggesting the absence of home-country bias in creating significant ties. Somewhat counterintuitively, Italian-Italian pairs are less likely to form significant ties compared to Italian-foreign and foreign-foreign pairs. 
When it comes to the fraction of relationship-dependent banks (Fig.~\ref{fig:Italian_foreign}d), the trend over the data period is similar to that of the fraction of Italian-Italian pairs (Fig.~\ref{fig:Italian_foreign}a). In particular, the percentage of Italian banks among all relationship-dependent banks is no less than 80\% throughout the data period. The deviation between the two lines in Fig.~\ref{fig:Italian_foreign}d suggests that the probability of becoming a relationship-dependent bank has been higher for Italian banks than for non-Italian banks at least until around 2012.

   \begin{figure}[t]
    \centering
     \includegraphics[width=.8\columnwidth]{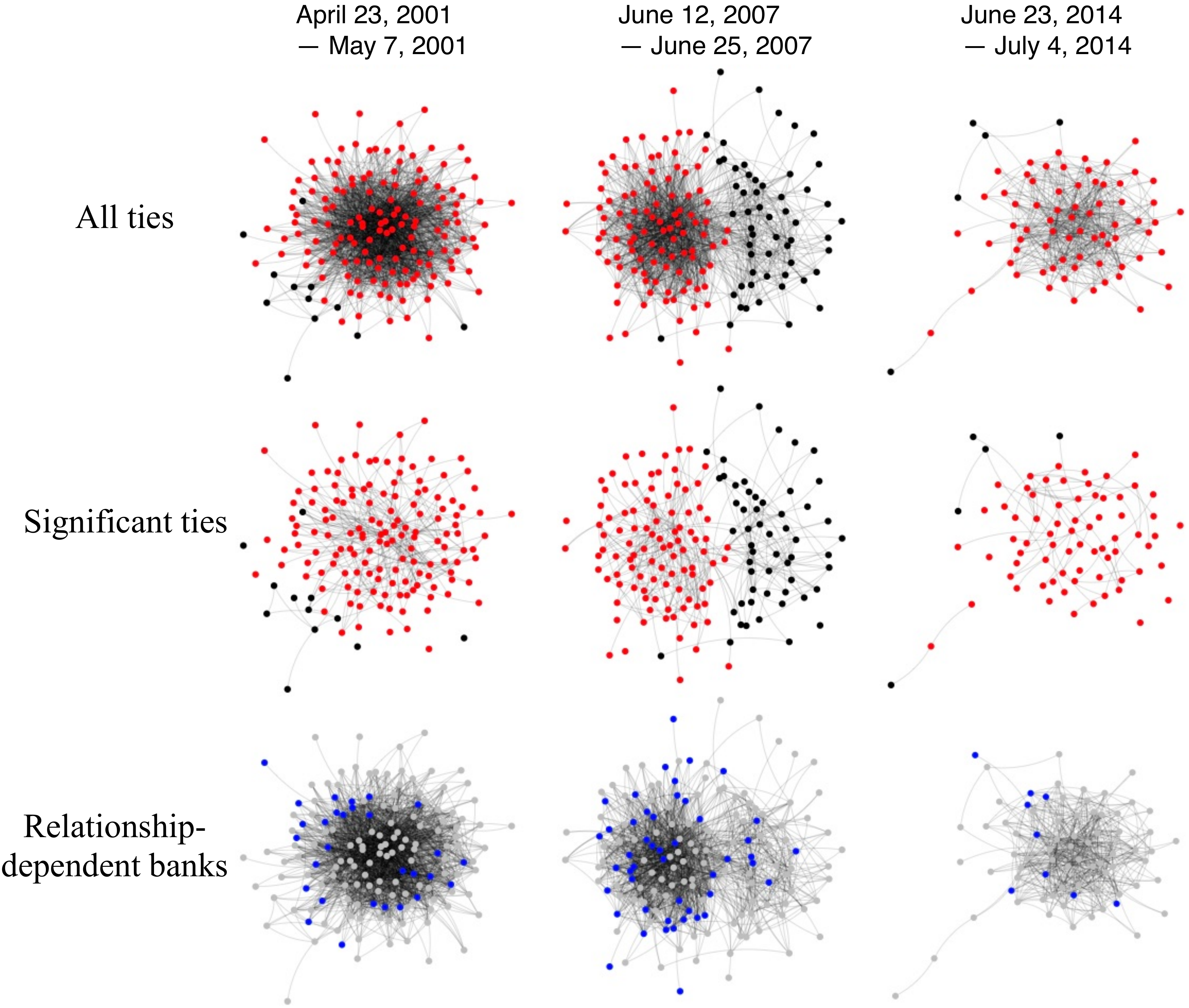}
    \caption{Visualization of aggregate networks. In the upper and middle rows, red and black circles represent Italian and foreign banks, respectively. In the bottom row, relationship-dependent banks are denoted by blue circles. The visualization is done by igraph package for Python ({\ttfamily{http://igraph.org/python/}}), using the Kamada-Kawai algorithm~\citep{Kamada1989algorithm}.
     }
    \label{fig:pref_net}
    \end{figure}

 Figure~\ref{fig:pref_net} presents a visualization of networks observed in different aggregate periods. 
In the early 2000s, there is no clear cut of groups since most active banks are Italian and they are well connected to each other. We observe a similar situation when we construct a network of significant ties only. By contrast, in a period shortly before the financial crisis, apparently there exist two tightly connected groups of banks, one formed by Italian banks and the other by foreign banks. This observation is explained by the result shown in Fig.~\ref{fig:Italian_foreign}; the fraction of foreign-foreign pairs reached its peak in 2007 while the fraction of Italian-foreign ties began to decrease in 2006. The two groups can be seen more clearly if we leave significant ties only since just a few significant ties connect Italian and foreign banks in this period. In 2014, the network looks similar to that in 2001, but the numbers of active banks and edges are much smaller in 2014 than in 2001.
In addition, the share of relationship-dependent foreign banks is relatively larger in the period during the crisis than in the pre- and post-crisis periods, although the vast majority of relationship-dependent banks are still Italian banks.

\subsection{Role of relationship lending}

The previous sections confirmed the existence of significant ties in the empirical data. In this section, we explore the difference in the outcomes of significant and non-significant ties in terms of their duration, trading conditions, and structural characteristics.

\subsubsection{Duration and the value of partnership}

\begin{figure}[t]
    \centering
     \includegraphics[width=.85\columnwidth]{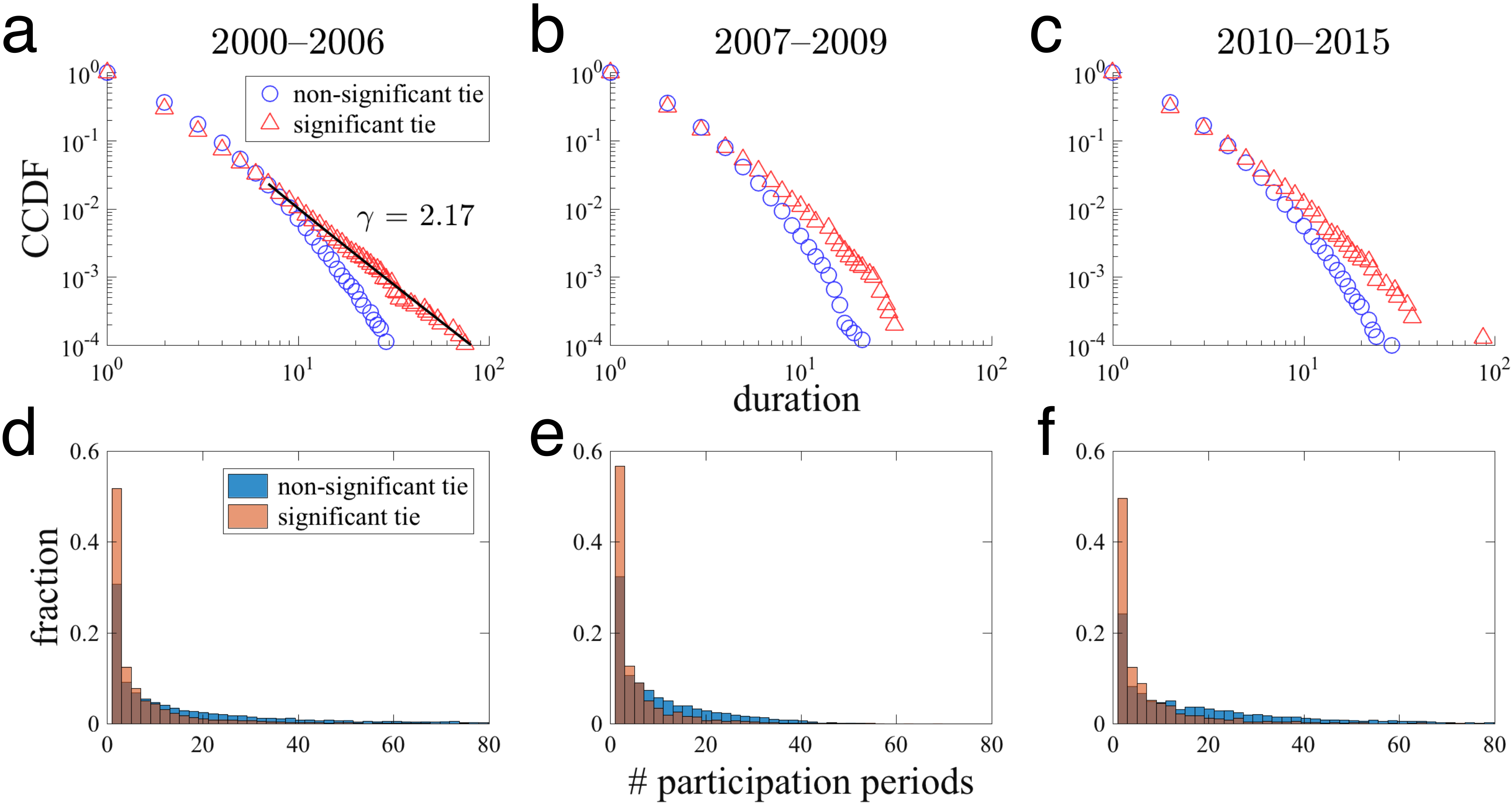}
    \caption{Duration of a lending relationship. (a)--(c): triangle (circle) denotes the complementary cumulative distribution function (CCDF) of the length of consecutive periods in each of which a bank pair is connected by a significant tie (a non-significant tie). In panel (a), the slope of the CCDF is also shown (black solid), which is estimated by the maximum-likelihood method proposed by \cite{Clauset2009Siam}. (d)--(f): histogram of the total number of periods in which a bank pair is connected by a significant tie or a non-significant tie.}
    \label{fig:duration}
    \end{figure}

 If relationship lending is understood as a long-lasting relationship between banks, the duration of significant ties should be longer than that of non-significant ties. 
Here, the duration of a (non-)significant tie between two banks is defined as the length of consecutive periods in each of which these banks form a (non-)significant tie between them.
In fact, the duration distribution of significant ties has a fatter tail than that of non-significant ties (Fig.~\ref{fig:duration}). The duration distribution of significant ties has a long tail and follows a power law at least in the pre-crisis period (2000--2006). 
This fat-tail behavior indicates that the longer the duration length, the more likely the current partnership will continue
 (i.e., the hazard rate is decreasing).  
To see this, let $P(d) = 1- (\kappa/\gamma) d^{-\gamma}$ $(\kappa >0)$ be a continuous approximation of the CDF of duration length $d$. The hazard rate $\lambda$, or the probability that a bank pair terminates their $d-$period relationship,  leads to 
\begin{align}
 \lambda (d) = \frac{p(d)}{1-P(d)} = \frac{\gamma}{d}, 
\end{align}
where $p(d)$ is the probability density function of $d$. It follows that during the pre-crisis period, the hazard rate at duration length $d$ is given by $\lambda(d) \approx 2.17 \: d^{-1}$. 

The decreasing hazard contrasts with the previous result for bank--firm relationships shown by \cite{Ongena2001JFE}. They found that the probability of terminating a relationship increases in duration, arguing that the value of relationships decreases over time. Our result indicates that the opposite holds true for the interbank market; the value of interbank relationships may increase over time. This is consistent with the traditional theory of relationship lending that supports the benefit of a long-term relationship~\citep{Freixas2008book}, suggesting that the longer the duration of a partnership, the greater the extent of private information owned by a lender~\citep{Sharpe1990JF}.

One might argue that the long duration of significant ties simply comes from the fact that relationship pairs tend to trade more frequently than non-relationship pairs do. However, Fig.~\ref{fig:duration}d--f reveals that the number of periods in which non-relationship pairs trade is larger than that of relationship pairs. Thus, the long duration of a significant tie is not attributed to the high frequency of the pair's trades.

\subsubsection{Terms of trades and the substitutability of trading partners}

In this section, we analyze the impact that the presence of a significant tie has on trade conditions (i.e., interest rates and the amount of loans).
To control for the influences of shifts in the policy rate and variations in the trading volume, we define the weighted average of detrended interest rates on bilateral transactions between banks $i$ and $j$ as
\begin{align}
 r_{t^{\prime},ij} \equiv \frac{\sum_{t\in D_{t^{\prime}}} {(r_{t,ij}^{\rm raw} - \langle r_t \rangle} ) w_{t,ij}}{\sum_{t\in D_{t^{\prime}}}w_{t,ij}},
\end{align}
where
\begin{align}
 \langle r_t \rangle & \equiv \frac{\sum_{i<j}r_{t,ij}^{\rm raw} \:w_{t,ij}}{\sum_{i<j}w_{t,ij}}, \label{eq:rbar}
\end{align}
and $r_{t,ij}^{\rm raw}$ is the raw interest rate. $w_{t,ij}$ is the total volume of funds traded between banks $i$ and $j$ on day $t$. Set $D_{t^{\prime}}$ represents the set of dates $t$ that belong to aggregate period $t^{\prime}$.
The average amount of loans per trade between banks $i$ and $j$ is defined as
\begin{align}
\overline{W}_{t^{\prime},ij} \equiv \sum_{t \in D_{t^{\prime}}} \frac{w_{t, ij}}{m_{t^{\prime},ij}},
\end{align}
where $m_{t^{\prime},ij}$ denotes the total number of trades between banks $i$ and $j$ during period $t^{\prime}$.

  \begin{figure}[t]
    \centering
     \includegraphics[width=.95\columnwidth]{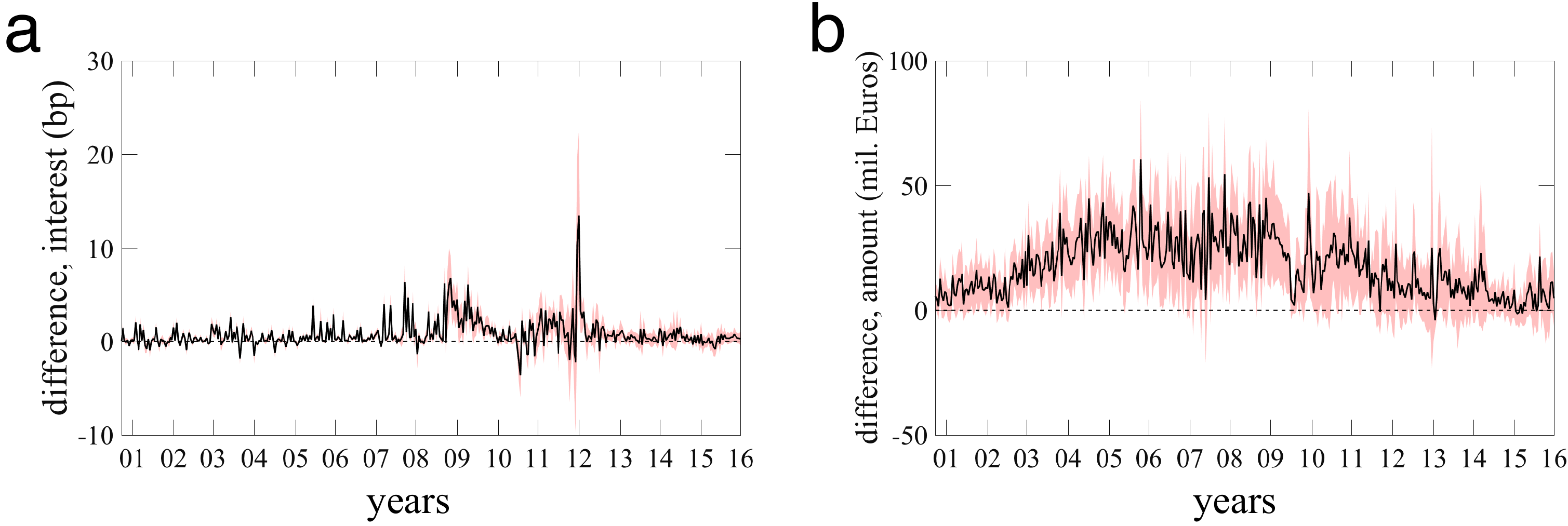}
    \caption{Impact of relationship lending on (a) interest rates and (b) trade amount. Solid line and shading indicate the average and the 95\% confidence interval, respectively. In each panel, the difference is calculated by subtracting the values for non-significant ties from those for significant ties.}
    \label{fig:diff_rate_amount}    
    \end{figure}

Figure~\ref{fig:diff_rate_amount} shows the differences in $r_{t^{\prime},ij}$ and $\overline{W}_{t^{\prime},ij}$ between significant ties and non-significant ties, calculated by subtracting the values for non-significant ties from those for significant ties.
The weighted interest rates are higher for relationship trades than for transactional trades by around three to six basis points during the global financial crisis. 
This fact implies the presence of imperfect substitutability of trading partners and that relationship lending played an important role in the management of liquidity~\citep{Affinito2012JBF}.\footnote{A price discrimination could occur if the maturity structures were different between relationship and transactional trades, but in this work we focus only on overnight transactions.}
 In interbank markets, it is occasionally observed that banks trying to meet urgent liquidity needs accept high interest rates to avoid stigma even if they can borrow from the central bank at lower rates~\citep{Ashcraft2011JMCB,Ennis2013RED}. The result shown in Fig.~\ref{fig:diff_rate_amount}a implies that those banks that played a role as ``lenders of last resort'' were connected with their borrowers by significant ties.

The upward spike in the difference in interest rates observed around January 2012 is considered to be caused by a ``longer-term refinancing operation (LTRO)'' introduced by the ECB. As pointed out by \cite{barucca2018organization}, the introduction of LTRO suddenly reduced the number of active banks and the volume of loans in the e-MID market. The decrease in the number of active banks might have undermined the substitutability of trading partners by limiting the number of potential partners, leading to an increase in the price of loans for relationship-dependent banks. 

\new{
\subsection{Extension}

 Here, we provide two extended analyses of significant ties. One is the analysis of a trading relationship among multiple banks. Since we can identify the significance of relationships between any combination of two banks, it is possible to investigate how likely a structure of direct transactions involving multiple banks connected by significant ties, such as a triangle, will emerge. Another extension considered here is the application of the identification of significant and non-significant ties to characterize the intraday behavior of banks.
 
\subsubsection{Relationship among multiple banks}
\label{sec:trilateral}
}

 In the literature of social network analysis, it has been widely recognized that there is a tendency that ``friends of friends are friends''~\citep{Wasserman1994book}.
This is called a \emph{triadic closure} since the two individuals having a friend in common often close the triangle~\citep{Easley2010book}.
Many studies have revealed that triadic closure plays an important role in achieving social cooperation~\citep{Hanaki2007ManSci}, determining the spread of a behavior across ties~\citep{Centola2010Science}, and understanding the long-term evolution of network structure~\citep{Lewis2012PNAS}, to name a few.
Analogously, the purpose of this section is to see whether triadic closures are also ubiquitous in the ``friendship'' network of banks. 
To be more precise, the question we address here is whether a significant tie is more likely to close a triangle of trading relationships (i.e., trilateral relationship) than a non-significant tie, provided that the triangle has at least two significant ties. This question is motivated by the well-known fact that triangles in social networks are mostly made of three strong ties~\citep{Granovetter1973AJS,Onnela2007PNAS,Easley2010book}. If a significant tie is more likely to close a triangle, it would indicate a previously unknown similarity between financial and social networks. In contrast, if a non-significant tie is more likely to close a triangle, then it would shed light on a unique characteristic of financial networks. 

\begin{figure}[t]
    \centering
     \includegraphics[width=.95\columnwidth]{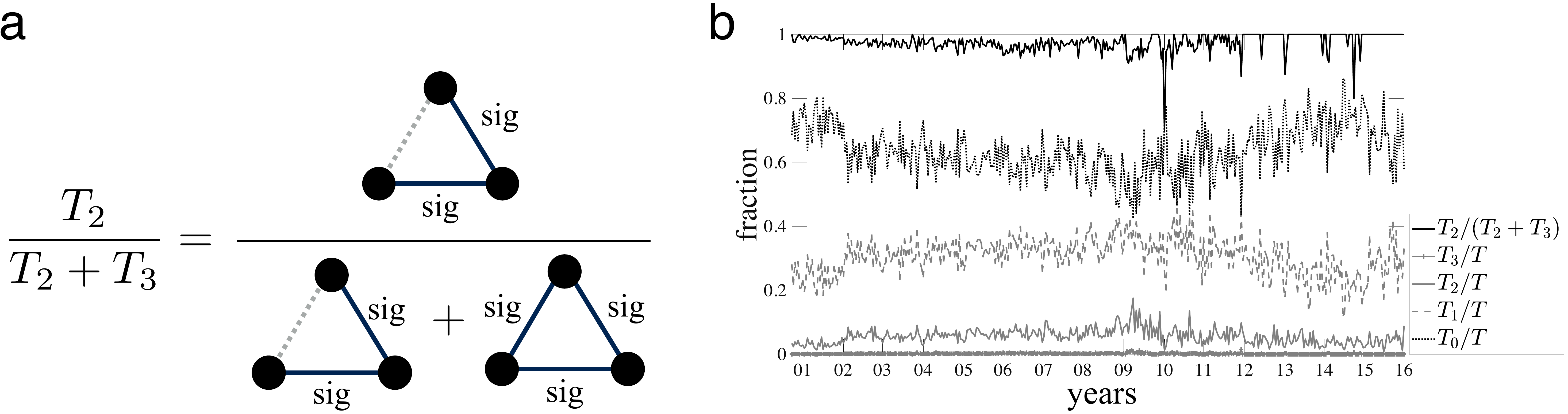}
    \caption{Trilateral relationship and significant ties. (a) Schematic of the probability $P_{\rm nonsig}  \equiv T_2/(T_2 +T_3)$ that a triangle having two significant ties has a non-significant closing tie. Solid edges labeled ``sig'' represent significant ties while dotted lines denote non-significant ties. (b) Time series of $P_{\rm nonsig}$ and the fraction of $\left\{ T_\ell \right\}_{\ell = 0,1,2,3}$ among all triangles.
 }
    \label{fig:triangle}
    \end{figure}

To answer this question, we first need to count the numbers of triangles in the aggregate networks having different numbers of significant ties (see Appendix for the procedure of calculation).
Let $T_\ell$ denote the number of triangles having $\ell$ significant ties $(\ell = 0, 1,2,3)$ in an aggregate network. The quantity we want to compute is schematically visualized in Fig.~\ref{fig:triangle}a; if $P_{\rm nonsig} \equiv T_{2}/(T_{2}+T_{3})$ is significantly larger than the fraction of non-significant ties in the whole network (i.e., the probability of placing a non-significant tie by chance), then the closing tie of a trilateral relationship is more likely to be a non-significant tie than random chance would suggest.
Since the percentage of significant ties is roughly 20\%--30\% throughout the data period (Fig.~\ref{fig:frac_pref}), the fraction of non-significant ties, denoted by $S_{\rm nonsig}\equiv |I_{\rm nonsig}|/\sum_{i<j}A_{ij}$, where $I_{\rm nonsig}$ is the set of non-significant ties, turns out to be around $0.7 - 0.8$, which becomes the baseline for evaluating $P_{\rm nonsig}$.

Figure~\ref{fig:triangle}b shows that $P_{\rm nonsig}$ is always above $0.9$ except for a few aggregate periods, meaning that a trilateral relationship having at least two significant ties tends to have a non-significant tie as the closing tie. This observation is statistically verified by the $t$-test for the null hypothesis that the means of $P_{\rm nonsig}$ and $S_{\rm nonsig}$ are equal, which is rejected with $p$-value $< 0.001$.
Figure~\ref{fig:triangle}b also illustrates the time series of $\left\{ T_\ell \right\}_{\ell = 0,1,2,3}$ normalized by $T$, the total number of triangles in each period.
The order $T_0 > T_1 > T_2 > T_3$ consistently holds true throughout the data period.
In addition, we see some trends in their relative shares; the share of $T_0$ and $T_1$ roughly move in opposite directions while the shares of $T_2$ and $T_3$ remain stable.

The result suggests that the local dynamics of tie formation in financial networks is quite different from that in social networks. While triangles of three strong ties are ubiquitous in networks formed by human interactions, interbank networks do not exhibit such a property.

\subsubsection{Intraday analysis}

     \begin{sidewaysfigure}
    \centering
     \includegraphics[width=.95\columnwidth]{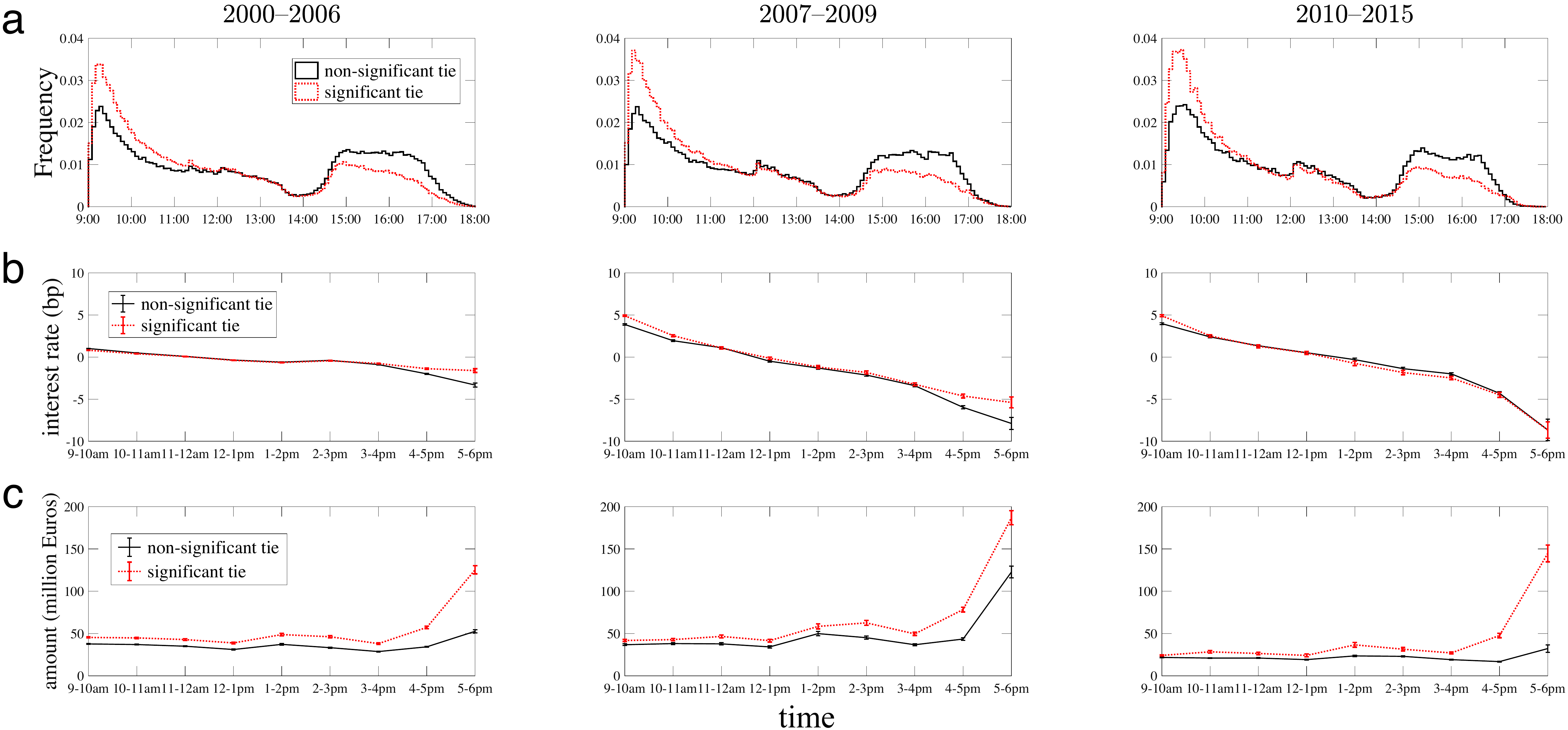}
    \caption{Impact of relationship lending on intraday interest rate and trade amount. In each panel, dotted and solid lines denote significant ties and non-significant ties, respectively. (a) Frequency of trades in each time interval.  (b) Mean deviation of interest rates from the daily mean.  (c) Average amount of trade. In panels (a) and (b), the error bar indicates the 95\% confidence interval.}
    \label{fig:timing_rate_amount}
    \end{sidewaysfigure}

 In the previous sections, we observed that bank pairs connected by significant ties exhibit different behaviors than other transactional pairs at a $\tau$-day aggregate scale. In this section, we explore intraday trading patterns to see if the existence of a significant tie has any impact on trades at higher frequencies.

In Fig.~\ref{fig:timing_rate_amount}, we observe subtle differences in the timing of intraday trading. A bank pair engaging in relationship lending tends to conduct a larger fraction of trades at early hours (9:00--11:00) and a smaller fraction of trades after 15:00 than a bank pair engaging in transactional trading (Fig.~\ref{fig:timing_rate_amount}a). This difference in the timing of trades does not seem to have a considerable impact on interest rates, but late-hour relationship trades resulted in slightly higher interest rates than those of transactional trades until the crisis period (Fig.~\ref{fig:timing_rate_amount}b).\footnote{The interest rate on the trade between banks $i$ and $j$ at time $\theta$ on day $t$ is defined as $r_{\theta,t,ij} = r_{\theta,t,ij}^{\rm raw} - \langle r_t \rangle$, where the superscript ``raw'' denotes the raw interest rate (before detrending) and $\langle r_t \rangle$ is defined in Eq.~\eqref{eq:rbar}.} 
Nevertheless, we still see a downward sloping term structure of intraday interest rates, which has been reported previously~\citep{Baglioni2010EL,Baglioni2008JMCB,Abbassi2017JMCB}.  

It is evident from Fig.~\ref{fig:timing_rate_amount}c that the positive difference in the trade amount between relationship and transactional lending tends to get larger as the market-closing time approaches. These gaps in the interest rate and amount of trades may reflect the fact that those banks that must obtain or release liquidity at the end of the market tend to rely on their partners to which they are connected by significant ties.

\subsection{Comparison with previous measures}\label{sec:comparison}

In closing this section, we assess the previously proposed measures of relationship lending by computing the extent to which they are able to detect significant ties.
 A naive measure of lending relationship is the frequency of interactions between two banks~\citep{Furfine1999microstructure,Kysucky2015MS,Brauning2017RF}:
\begin{align}
{\rm RL}_{ij,t^{\prime}} \equiv \log ( 1 + m_{ij,t^{\prime}}),
\label{eq:RL}
\end{align}
which denotes the logarithm of the number of transactions between banks $i$ and $j$ conducted in an aggregate period $t^{\prime}$. 
 The second and more widely used measure is the \emph{lender--preference index} (LPI)~\citep{Cocco2009JFI,Affinito2012JBF,Craig2015JBF,Brauning2017RF}:  
\begin{align}
{\rm LPI}_{ij,t^{\prime}} \equiv \frac{\sum_{t\in D_{t^{\prime}}}w_{ij,t}}{\sum_{j:j\neq i}\sum_{t\in D_{t^{\prime}}}w_{ij,t}}.
\label{eq:LPI}
\end{align}
${\rm LPI}$ captures the degree of concentration of lending to a particular partner.\footnote{Note that the amount lent is equal to the amount borrowed in our setting since we only treat undirected networks.}  If the fraction of funds lent to a particular partner is high, then it would indicate the existence of relationship lending. These two conventional measures are usually employed as explanatory variables of linear regression models.

\begin{figure}[t]
    \centering
    \includegraphics[width=.6\columnwidth]{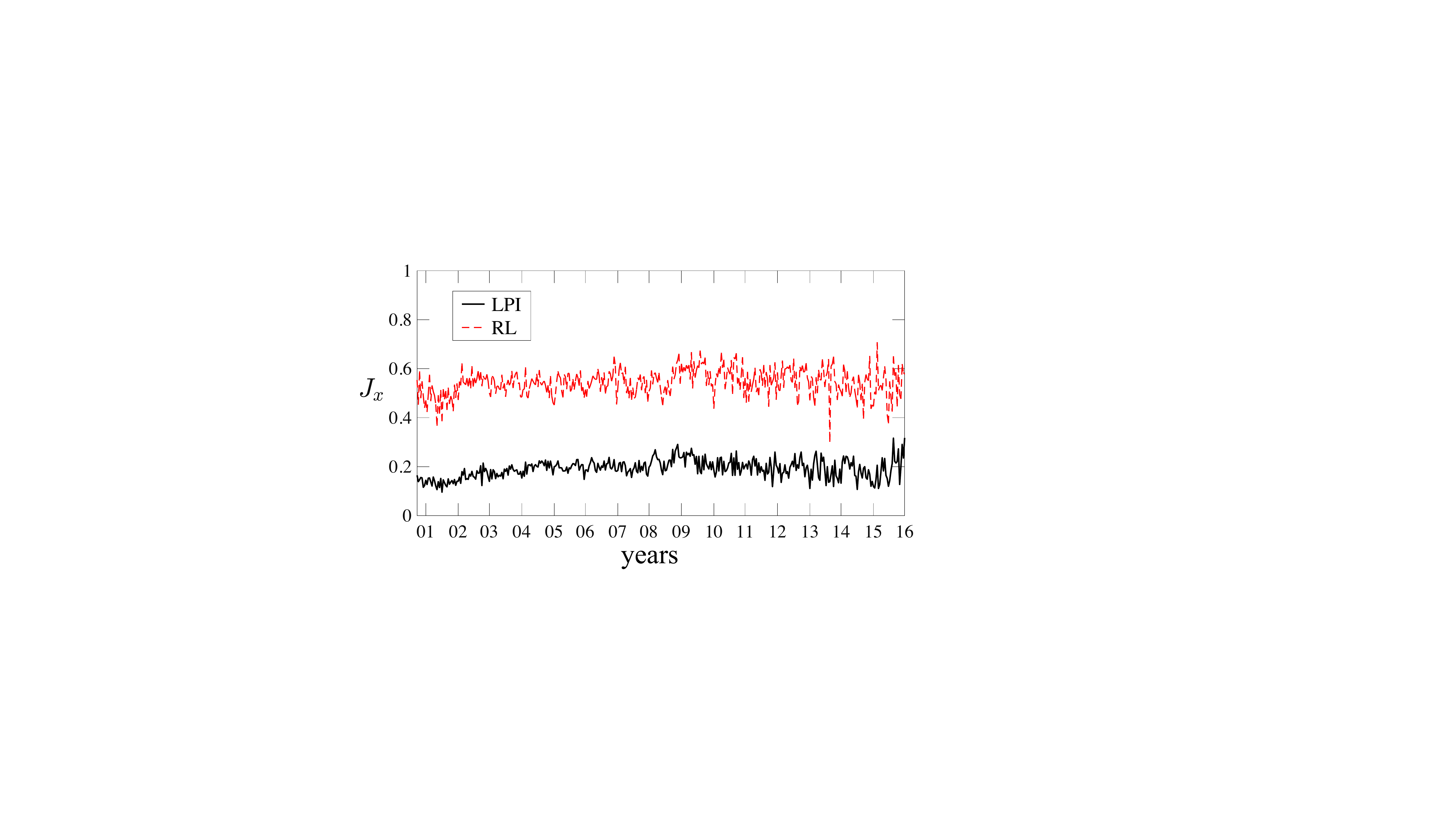}
  \caption{Comparison with the conventional measures of relationship lending. Solid (dashed) line represents the ``accuracy'' of LPI (RL), defined by the Jaccard index (Eq.~\eqref{eq:jaccard}). ${\rm RL}$ and ${\rm LPI}$ are given by Eqs.~\eqref{eq:RL} and \eqref{eq:LPI}, respectively.  }
    \label{fig:LPI}
    \end{figure}

  Now let us assess the accuracy of $\RL$ and $\LPI$ in terms of their detectability of significant ties. Let $I_{\rm sig}$ denote the set of significant ties identified by our edge-based test, which is treated as the ``ground truth.'' The fraction of significant ties among all ties in a given period is represented as $S_{\rm sig} \equiv |I_{\rm sig}|/\sum_{i<j}A_{ij}$.
 For each $x=\RL,\LPI$, let $I_{x}$ be the set of bank pairs whose score of $x$ is ranked top $S_{\rm sig}\%$ in the corresponding period. If measure $x$ correctly reflects the strength of a bilateral relationship, then the following Jaccard index will take a value close to one:
 \begin{align}
      J_{x} = \frac{|I_{x}\cap I_{\rm sig}|}{|I_{x}\cup I_{\rm sig}|}, \;\; x=\RL, \LPI.
      \label{eq:jaccard}
 \end{align}         
 
  Figure~\ref{fig:LPI} reveals that, somewhat surprisingly, $\RL$ outperforms $\LPI$, although the definition of $\LPI$ seems more sophisticated than that of $\RL$ as a measure of a lending relationship. 
This may be due to the fact that the degree of concentration of lending to or borrowing from a particular bank does not necessarily relate to the number of trades with the bank, whereas our definition of a significant tie generally favors a bank pair conducting a large number of trades.
$\RL$ just reflects the number of bilateral trades regardless of their volume, but it is closer to our idea of significant ties than $\LPI$ is. Of course, just counting the number of trades is not enough because one must take into account the difference in banks' activity levels. An observation of repeated trades between two banks does not necessarily lead to the presence of a significant tie because such repetitive trades may be explained by random chance if the two banks exhibit high activity levels. Nevertheless, Fig.~\ref{fig:LPI} shows that $\RL$ is far more appropriate than $\LPI$ as a measure of relationship lending, although $J_{\RL} \sim 0.6$ does not mean that $\RL$ is very accurate.

 \new{
\section{Robustness check}

 In this section, the robustness of the baseline framework is investigated. We first examine the power of the proposed test on synthetic core-periphery networks. We also perform edge- and node-based tests in a more general setting where bank activity is time-varying and/or edges are directed.       

\subsection{Monte Carlo analysis with core-periphery structure}

 We check the power of the proposed test by numerical simulation. To generate a sequence of synthetic daily networks on which significance tests are based, we employ a core-periphery structure since it has been shown to be a plausible network structure in various interbank markets~\citep{Imakubo2010BOJ,craig2014interbank,Fricke2015CompEcon}.
 
 The procedure of the Monte Carlo analysis is as follows:
\begin{enumerate}
\item Initially there are $N$ isolated banks. Fraction $f_{\rm c}$ of the banks are designated as core banks and fraction $1-f_{\rm c}$ as peripheral banks.   
\item On day $t$, any two core banks are connected with probability $p_{\rm cc}$, a core bank and a peripheral bank are connected with $p_{\rm cp}$, and there is no edge between two peripheral banks. We generate a sequence of $\tau$ snapshots of daily interbank networks, $\widehat{A}(t),\widehat{A}(t+1),\ldots,\widehat{A}(t+\tau)$.
\item Among the pairs that had at least one transaction within $\tau$ days, choose a fraction $f_{\rm rel}$ of pairs at random as relationship pairs. For a relationship pair $(i,j)$, assume that the probability that an additional trade is not imposed at $t$, denoted by $p_{ij}^{\rm norel}(t)$, depends on the number of consecutive trading days up to $t-1$. The hazard rate is given by
\begin{align}
 p_{ij}^{\rm norel}(t) = \frac{b_{0}}{b_{1}+b_{2}D_{ij}(t-1)},
 \label{eq:relprob}
\end{align}
where $D_{ij}(t-1)$ denotes the number of consecutive transactions between $i$ and $j$ up to $t-1$, and $b_{0}$, $b_{1}$ and $b_{2}$ are non-negative parameters.  
\item If $\widehat A_{ij}(t)=0$, add a relationship edge $(i,j)$ with probability $1-p_{ij}^{\rm norel}(t)$ for all relationship pairs $(i,j), i \neq j$. This gives us the sequence of adjacency matrices with relationship edges, $\{\widehat{A}_{\rm rel}(t)\}$.
\item Estimate bank activity $\{a_{i}\}$ using $\{\widehat{A}_{\rm rel}(t)\}$ in the same way as described in section~\ref{sec:MLE} and implement the edge-based tests.
\end{enumerate}

It should be noted that if there is no relationship edge in the synthetic networks (i.e., networks $\{\widehat{A}(t)\}$), then the number of transactions between two banks follows a binomial distribution since in each day a bilateral edge $(i,j)$ is created with a constant probability $u(a_i,a_j)$. By contrast, if two banks are matched in a non-random manner, then the number of connections no longer obeys a binomial distribution. In the latter case, the presence of non-random edges should be detected by the proposed tests.

We run simulations 5,000 times with the length of simulation periods 3,000. For significance tests, only the last $\tau$ periods are used and the initial $(3000-\tau)$ periods are discarded. The parameter values are set as follows: $f_{\rm c}=0.5$, $p_{\rm pp}=0.06$, $p_{\rm cp}=0.03$, $b_{0}=1$ and $f_{\rm rel}=0.2$. We check different values of $b_{1}$ and $b_{2}$.

\begin{figure}[t]
    \centering
    \includegraphics[width=.85\columnwidth]{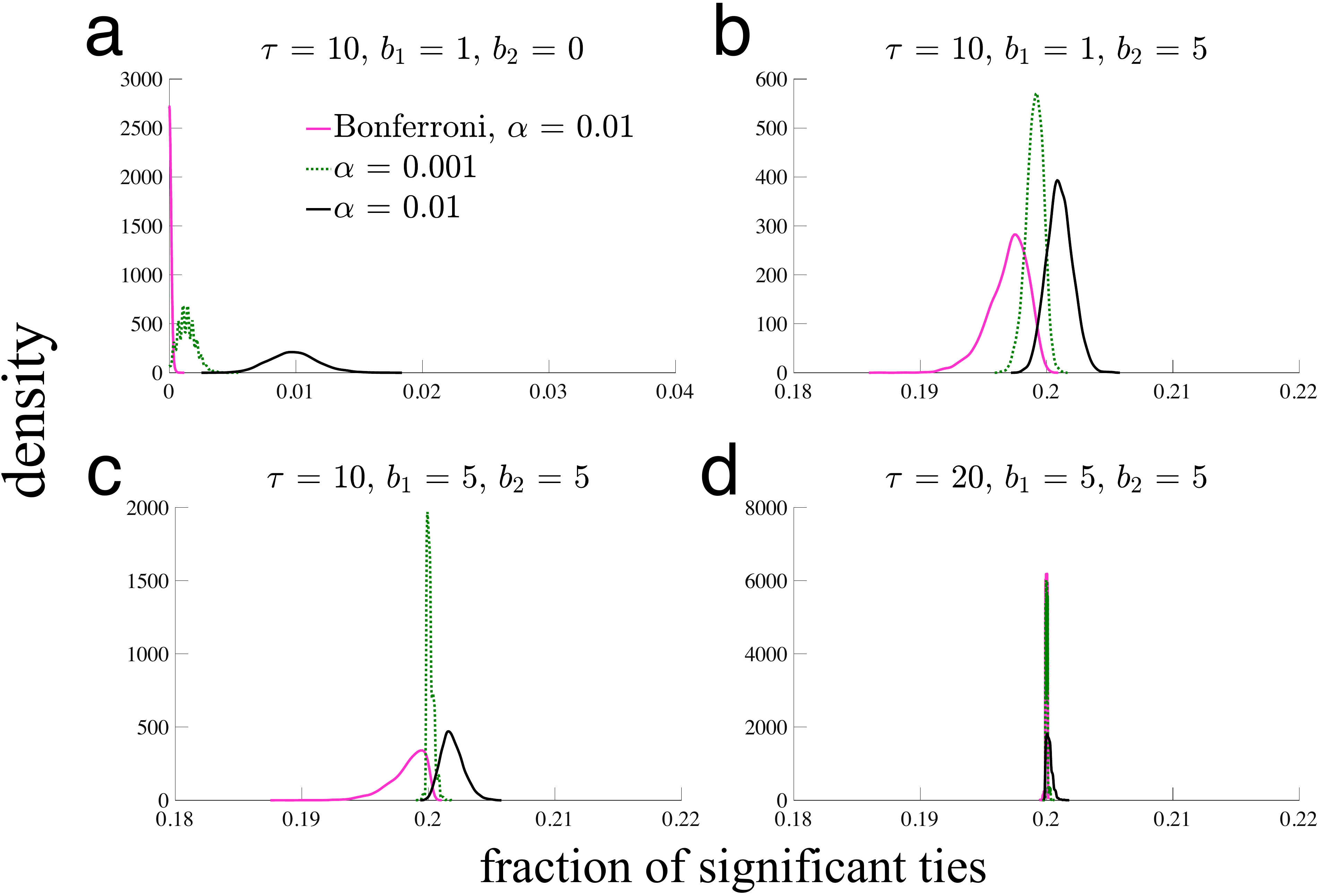}
  \caption{\new{Edge test on synthetic networks with core-periphery structure. The true fraction of significant ties is set at 0.2. $\alpha$ denotes the significance level.}}
    \label{fig:synthetic}
    \end{figure}

Fig.~\ref{fig:synthetic} illustrates the density functions of the fraction of detected significant ties. 
Fig.~\ref{fig:synthetic}a corresponds to the case of no relationship lending (i.e., $b_2 =0$), in which the number of transactions between two banks follows a binomial distribution. We see type-I errors in Fig.~\ref{fig:synthetic}a because multiple tests are implemented, but the tests with Bonferroni correction alleviates the problem. Fig.~\ref{fig:synthetic}b and c introduce relationship lending into the otherwise random network with a core-periphery structure. It turns out that the proposed tests are able to detect significant ties quite accurately as long as there is a certain extent of non-random relationship. Fig.~\ref{fig:synthetic}d shows that increasing the length of time window, $\tau$, may improve the accuracy of the tests on synthetic networks.

\subsection{Different time windows}

  In the baseline framework, we split the whole data period into non-overlapping $t_{\rm max}^{\prime} = \lfloor t_{\rm max}/\tau \rfloor$ time windows, each of which consisting of $\tau$ business days. To check the sensitivity of the results to the way we split the data, we implement significance tests by using rolling time windows for different values of $\tau$.

  Fig.~\ref{fig:frac_pref_robustcheck} presents the results for $\tau=\{5,10,20\}$, in which we progressively slide the start date of a $\tau$-day time window by one day increments. We see that introducing rolling time windows does not have a quantitative impact on the fractions of significant ties and relationship-dependent nodes.
   The figure also indicates that time windows of $\tau=5$ may be too narrow to capture banks' relationship dependency since it gives us much lower fractions of relationship-dependent banks compared to the cases of $\tau=10$ and $20$. On the other hand, the results for $\tau=10$ and $20$ are quite similar, which suggests that the choice of $\tau=10$ would be appropriate given the fact that an increase in $\tau$ also has negative effects on the accuracy of the Poisson approximation while improving the consistency of maximum-likelihood estimates (Fig.~\ref{fig:synthetic}).

\subsection{Time-varying bank activity}\label{sec:time-varying}
 
  In the baseline null model, we assumed that activity level $\vect{a}$ is constant within a time interval. Here, we relax this assumption by allowing $\vect{a}$ to fluctuate at the daily scale. The matching probability between banks $i$ and $j$ is given by
  \begin{align}
      u(a_i(t),a_j(t)) \equiv a_i(t) a_j(t), \; \forall\: i,j,t,
  \end{align}
  where $a_i(t)$ denotes the activity of bank $i$ on day $t$. Thus, we need to estimate $N\times \tau$ activity parameters, $(\vect{a}(1),\ldots, \vect{a}(\tau))$. 
  
  \subsubsection{Edge-based test}
    The procedure for the edge-based test based on variable activities is as follows:
\begin{enumerate}
    \item By imposing $\tau=1$ in Eq.~\eqref{eq:ML_a}, we obtain the estimates of activities on day $t$, denoted by $\hat{\vect{a}}(t)$, by solving the following $N$ equations:
    \begin{align}
 \hat{H}_i(\hat{\vect{a}}(t)) \equiv & \sum_{j:j\neq i}\frac{A_{ij}(t) - \hat{a}_{i}(t)\hat{a}_{j}(t)}{1-\hat{a}_{i}(t)\hat{a}_{j}(t)} = 0, \; \forall \: i = 1,\ldots N,
 \label{eq:ML_avar} 
 \end{align}
where $A_{ij}(t)$ denotes the $(i,j)$th element of a binary adjacency matrix of day $t$.\footnote{\new{As is shown in Fig.~\ref{fig:var_const_activity}, the estimated daily activities fluctuate around the constant activity levels. }} 
 Under the null, the total number of transactions between banks $i$ and $j$ in a given time interval, denoted by $m_{ij}$, obeys a Poisson binomial distribution with mean $\hat{\lambda}_{ij}\equiv\sum_{t=1}^{\tau}u(\hat{a}_i(t),\hat{a}_j(t))$ and variance $\hat{\sigma}_{ij}\equiv\sum_{t=1}^{\tau}(1-u(\hat{a}_i(t),\hat{a}_j(t)))u(\hat{a}_i(t),\hat{a}_j(t))$.
     
    \item Approximate the Poisson binomial distribution of $m_{ij}$ by a Poisson distribution:
    \begin{align}
   f(m_{ij}|\{\hat{\vect{a}}(t)\}) \approx \frac{\hat{\lambda}_{ij}^{m_{ij}}e^{-{\hat{\lambda}_{ij}}}}{m_{ij}!} \equiv \widetilde{f}(m_{ij}|\{\hat{\vect{a}}(t)\}), \label{eq:poisson}
   \end{align} 
   where the error bound is given by the Le Cam's theorem:
   \begin{align}
       \sum_{m_{ij}=0}^\infty\left| f(m_{ij}|\{\hat{\vect{a}}(t)\}) - \frac{\hat{\lambda}_i^{m_{ij}}e^{-\hat{\lambda}_{ij}}}{m_{ij}!}\right| < \frac{2(1-e^{-\hat{\lambda}_{ij}})}{\hat{\lambda}_{ij}}\sum_{t=1}^{\tau}u(\hat{a}_i(t),\hat{{a}}_j(t))^2,  \; \forall \; i,j.
   \end{align}
   \item Implement the edge-based tests by using Eq.~\eqref{eq:poisson} as a null distribution.
   
  \end{enumerate}
    
     Fig.~\ref{fig:undirected_varact}a shows that the qualitative result does not change even after introducing variable activity parameters while the detected fraction of significant ties is slightly lower than before.   
    
    \begin{figure}[t]
    \centering
    \includegraphics[width=.85\columnwidth]{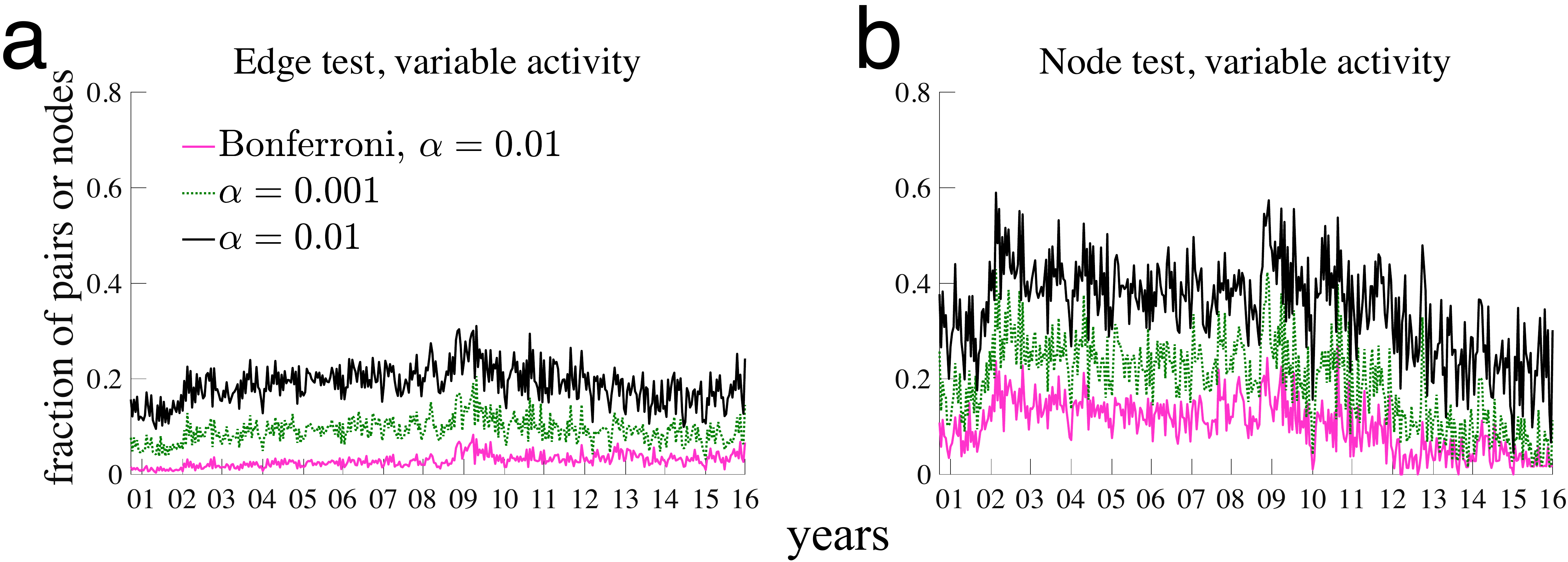}
  \caption{\new{Fractions of significant ties and relationship-dependent banks detected by a null model with variable activity.}}
    \label{fig:undirected_varact}
    \end{figure}

    \subsubsection{Node-based test}
    
    The only modification for the node-based test is that we now take into account the fact that the probability of matching between two nodes can change over time. Here, the probability that bank $i$ has at least one transaction with bank $j$ in a given period is given by $1-\prod_{t=1}^{\tau}(1-u(\hat{a}_{i}(t),\hat{a}_{j}(t)))$. Accordingly, aggregate degree $K_{i}$ (i.e., the number of bank $i$'s unique trading partners) follows a Poisson binomial distribution with mean $\hat{\lambda}_{i} \equiv \sum_{j:j\neq i}\left[1-\prod_{t=1}^{\tau}(1-u(\hat{a}_{i}(t),\hat{a}_{j}(t)))\right]$. The distribution of $K_i$ is then approximated by a Poisson distribution:
    \begin{align}
   f(K_i|\hat{\vect{a}}) \approx \frac{\hat{\lambda}_i^{K_i}e^{-{\hat{\lambda}_i}}}{K_{i}!} \equiv \widetilde{f}(K_i|\hat{\vect{a}}). 
\label{eq:df_poiss_var}
\end{align} 

Fig.~\ref{fig:undirected_varact}b indicates that again the introduction of variable activities does not change the time-series behavior of the fraction of relationship-dependent banks.

\subsection{Directed edges}\label{sec:directed}
 
 Our statistical tests can also incorporate the directionality of edges. Here, we need to consider two sorts of bank activities: in-activity and out-activity. 
 The random probability that bank $i$ lends to bank $j$ is now given by
\begin{align}
      u_{i\to j}(a_{i}^{\rm out},a_{j}^{\rm in}) = a_{i}^{\rm out}a_{j}^{\rm in},
      \label{eq:matching_dir}
\end{align}
where $a^{\rm in}$ and $a^{\rm out}$ denote in- and out-activity, respectively.
The maximum-likelihood estimates of in- and out-activity are the solution for the following $2N$ equations:
\begin{align}
 \sum_{j:j\neq i}\frac{m_{ij}-\tau a_{i}^{\rm out}a_{j}^{\rm in}}{1-a_{i}^{\rm out}a_{j}^{\rm in}} &= 0, \\
 \sum_{j:j\neq i}\frac{m_{ji}-\tau a_{j}^{\rm out}a_{i}^{\rm in}}{1-a_{j}^{\rm out}a_{i}^{\rm in}} &= 0,  
 \end{align}
for $i = 1,\ldots, N$. Using the matching probability Eq.~\eqref{eq:matching_dir} as a parameter, we can test the significance of a directed edge in the same way as explained in section~\ref{sec:edge_test}. The dependency of a node on particular creditors (i.e., \emph{borrowing dependency}) or borrowers (i.e., \emph{lending dependency}) can also be tested by implementing a directed version of the node-based test. 

 It is also straightforward to introduce a daily variations of activity (see section~\ref{sec:time-varying}) into the directed version of significance tests. The daily in- and out-activity on day $t$ are estimated by solving the following $2N$ equations: 
\begin{align}
 \sum_{j:j\neq i}\frac{A_{ij}(t)- a_{i}^{\rm out}(t)a_{j}^{\rm in}(t)}{1-a_{i}^{\rm out}(t)a_{j}^{\rm in}(t)} &= 0,\\ \sum_{j:j\neq i}\frac{A_{ji}(t)- a_{j}^{\rm out}(t)a_{i}^{\rm in}(t)}{1-a_{j}^{\rm out}(t)a_{i}^{\rm in}(t)} &= 0,\end{align}
for $ i = 1,\ldots, N$. Then we estimate $2N\times \tau$ parameters for each time window.

\begin{figure}[t]
    \centering
    \includegraphics[width=.99\columnwidth]{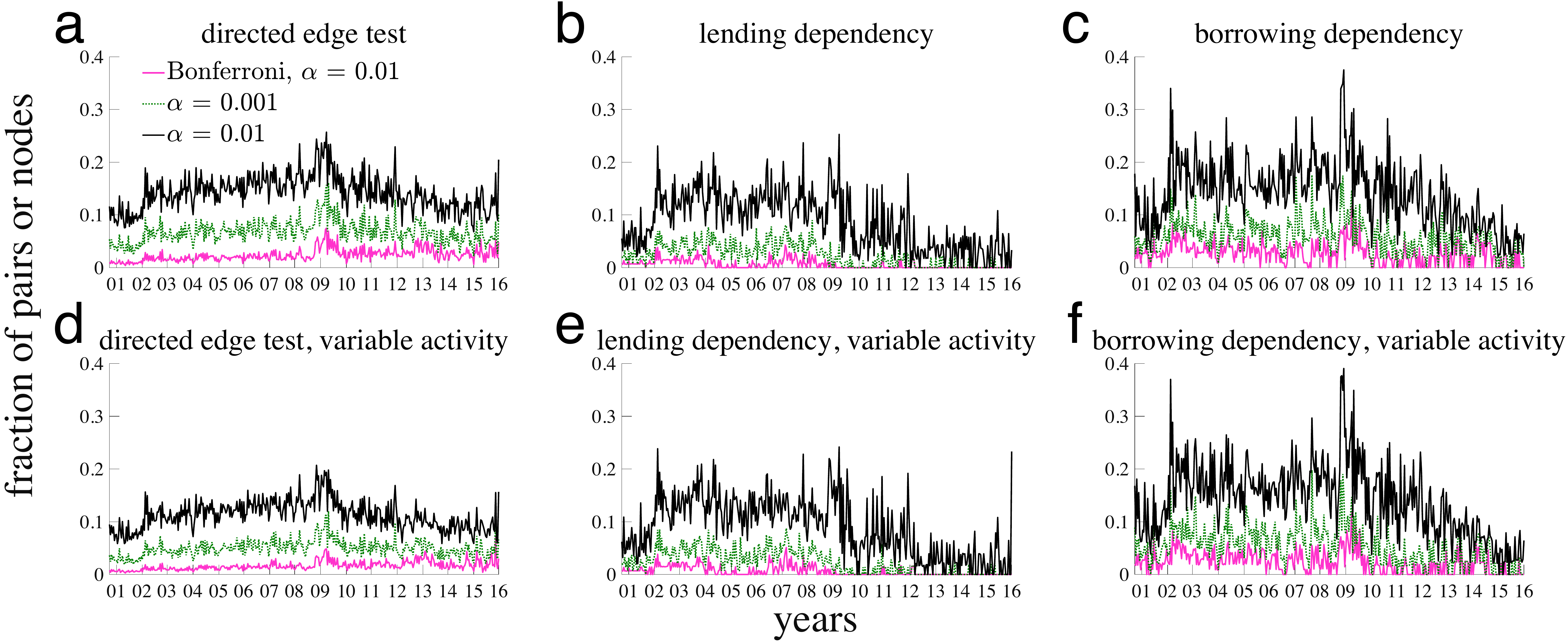}
  \caption{\new{Edge- and node-based tests for directed networks. Upper panels (i.e., a--c) represent the fraction of edges or banks that passed the tests for directed networks based on constant bank activity (see section~\ref{sec:directed}). Lower panels (i.e., d--f) show the corresponding results for the tests based on variable bank activity (see section~\ref{sec:time-varying}).}}
    \label{fig:directed_test}
    \end{figure}

The directed edge test in fact yields essentially the same result as the one we obtained in the undirected model (Fig.~\ref{fig:directed_test}a and d). A reason for this is that there are few pairs that have bidirectional edges~\citep{Kobayashi2017arxiv}. 

 On the other hand, we see an interesting property for the directed node-based tests. For a sufficiently high level of statistical significance, the fraction of borrowing-dependent banks spiked around the global crisis (Fig.~\ref{fig:directed_test}c and f) while the fraction of lending-dependent banks did not (Fig.~\ref{fig:directed_test}b and e).  This strongly suggests that the rise of relationship-dependent banks in the midst of the global financial crisis, as indicated by the undirected node test, could be attributed to the increased fraction of borrowing banks that relied on a limited number of creditors. 
}

\section{Conclusion and discussion}

This work proposed a statistical test for identifying bank pairs that are engaging in relationship lending by introducing the concept of a significant tie. The proposed identification test was applied to the Italian daily interbank networks formed by overnight transactions. The point of our identification method is that we test whether or not the number of trades between two banks can be explained by random chance after controlling for the intrinsic activity levels of those banks. If the number of trades is statistically significant (i.e., cannot be explained by random chance), then we say that the two banks are connected by a significant tie.
We showed that the percentage of significant ties among all ties has been quite stable over the past years, while the number of significant ties itself has been declining along with the total number of trades in the interbank market.  

We found several important properties that distinguish relationship lending from other transactional lending. First, the duration of a significant tie is, on average, longer than that of a non-significant tie. This property indicates that the value of continuing a relationship increases in duration, as suggested by many theoretical studies~\citep{Freixas2008book}.
Second, in the midst of financial distress, banks in need of liquidity relied on banks to which they are connected by significant ties even at the cost of high interest rates. This may be evidence that relationship lenders played a role as the ``lender of last resort" during financial turmoil. Third, there is no home-country bias in creating significant ties. 

While we apply the proposed identification method to the Italian interbank market due simply to data availability, in principle it would also be possible to implement it on various time-varying networks. 
Our method is quite general and thereby not limited to the use of identification of relationship lending in interbank markets. For example, they could also be applied to trading networks in the corporate bond market~\citep{DIMAGGIO2017JFE} and the municipal bond market~\citep{Li2014dealer} in search of a hidden structure of significant ties between market traders. 

The current work also provides a temporal-network analysis of the interbank market, which is still scarce in the field of network science and in economics, with a few exceptions~\citep{Kobayashi2017arxiv,barucca2018organization}. Understanding the dynamic formation of interbank networks is quite important because the network structure formed by overnight bilateral transactions drastically changes day to day, meaning that the risk of financial contagion varies over time. While most of the studies on financial systemic risk are based on static networks~\citep{GaiKapadia2010,Cont2013,Brummitt2015PRE}, in the real world the risk of financial contagion emerges on networks with time-varying structures. We hope that our work will advance our knowledge about the mechanism of temporal financial networks, which could contribute to the real-time management of financial stability. 　


\vspace{1cm}

\section*{Appendix:  Counting the number of triangles}

In section \ref{sec:trilateral}, we counted the number of triangles in an aggregate network to investigate the role of significant ties in a trilateral relationship. 
Computing the number of triangles having $k$ significant ties, $T_{k}$, is straightforward if we exploit the power of adjacency matrix. First, the total number of triangles in the whole network is given by $T = \sum_{k=0}^{3}T_{k} =  {\rm tr}{(A^{3})}/6$, where ${\rm tr(\cdot)}$ denotes trace. This equality is based on the fact that the $(i,j)$ element of $A^{n}$ represents the number of paths from $i$ to $j$ that can be reached at exactly $n$ steps. Therefore, the diagonal elements of $A^{3}$ contain the numbers of triangles. Second, the number of triangles formed by three significant ties is given as $T_{3} = {\rm tr}{(A_{\rm sig}^{3})}/6$, where $A_{\rm sig}$ is an adjacency matrix of the network consisting only of significant ties.  Third, the number of triangles formed only by non-signifiant ties leads to $T_{0} = {\rm tr}{((A-A_{\rm sig})^{3})}/6$. Fourth, the number of triangles having exactly one and two significant ties, $T_1$ and $T_2$, are obtained as follows:
 \begin{enumerate}
 \item Create a ``signed'' adjacency matrix $A_{\rm signed}$, where $(A_{{\rm signed}})_{ij} = 1$ if $i$ and $j$ are connected by a significant tie, $-1$ if connected by a non-significant tie, and 0 otherwise. 
 \item Compute $T_{\rm signed} = {\rm tr}{(A_{\rm signed}^{3})}/6$. This is equal to the difference between the number of triangles having an odd number of significant ties and the number of triangles having an even number of significant ties (i.e., $T_{\rm signed} = (T_{1}+T_{3}) - (T_{0}+T_{2}))$. 
  \item Derive $T_1$ and $T_{2}$ by substituting $T$, $T_{0}$, and $T_{3}$ into equations $T=\sum_{k=0}^{3}T_{k}$ and $T_{\rm signed} = (T_{1}+T_{3}) - (T_{0}+T_{2})$. 
 \end{enumerate}
This procedure gives us $T_{k}$ for $k=0,1,2,3$.

\section*{Acknowledgements}
T.K. acknowledges financial support from the Japan Society for the Promotion of Science Grants no.~15H05729 and 16K03551. T.T. was supported by JST ERATO Grant Number JPMJER1201, Japan.



%


\ifx\undefined\bysame
\newcommand{\bysame}{\hskip.3em \leavevmode\rule[.5ex]{3em}{.3pt}\hskip0.5em}
\fi

\clearpage
\onecolumn

\setcounter{section}{0}
\setcounter{table}{0}
\setcounter{equation}{0}
\setcounter{figure}{0}
\setcounter{page}{1}
     
\renewcommand{\thetable}{S\arabic{table}}
\renewcommand{\thefigure}{S\arabic{figure}}
\renewcommand{\thesection}{S\arabic{section}}
\renewcommand{\theequation}{S\arabic{equation}}


\fontsize{18pt}{22pt}\selectfont

{}\
\vspace{1cm}

\vspace{2cm}
\center{{\textbf{Supplementary Information:}}  \\
\vspace{.5cm}
\center{\textbf{ ``Identifying relationship lending in the interbank market: A network approach''}}} 

\vspace{1cm}
\fontsize{13pt}{0pt}\selectfont

\fontsize{9pt}{11pt}\selectfont

  \begin{figure}[b]
    \centering
       \includegraphics[width=.9\columnwidth]{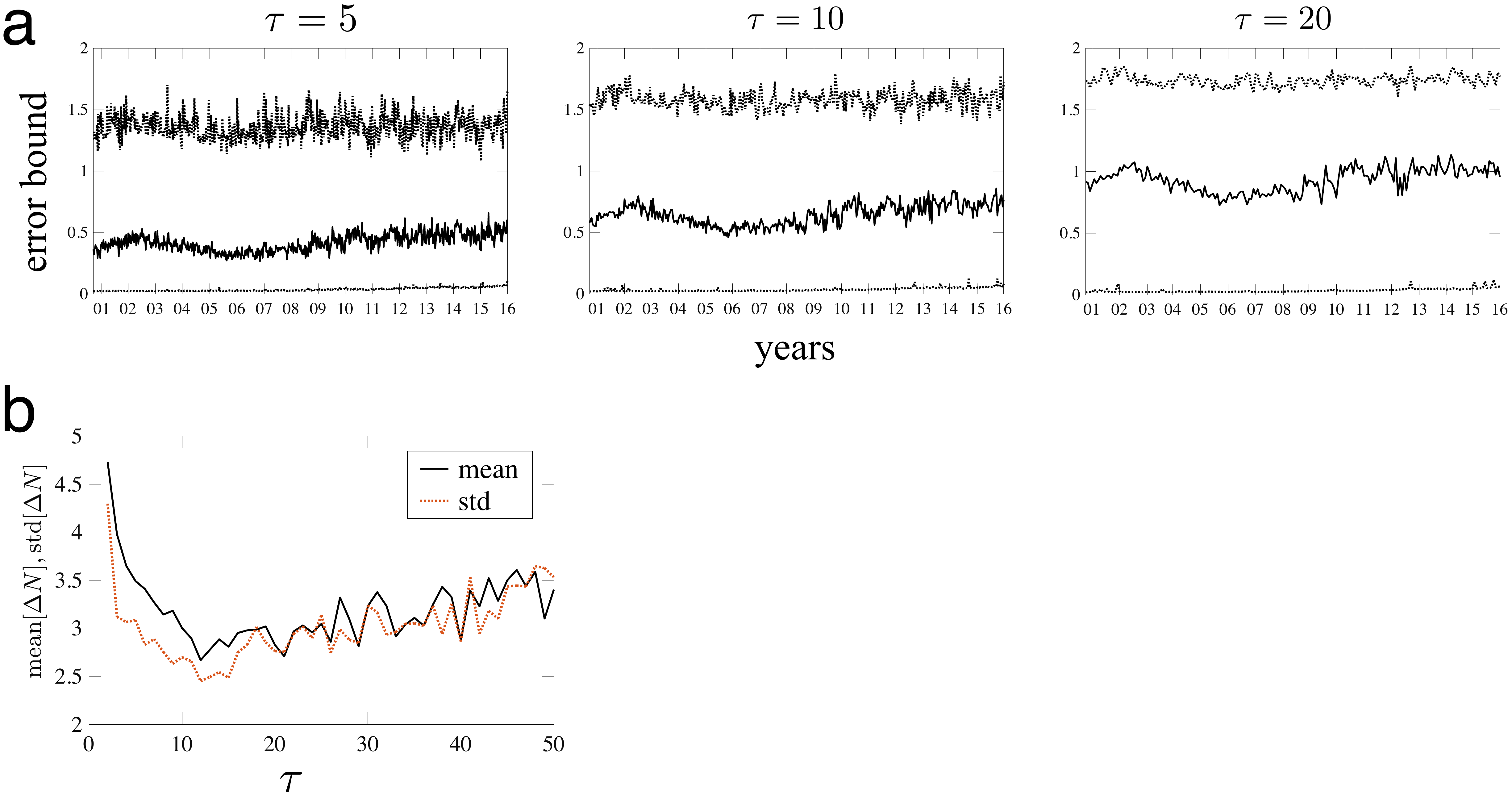}
           \caption{Effects of varying aggregate length $\tau$. (a) An increase in $\tau$ makes it less accurate the Poisson approximation of a Poisson binomial distribution. Solid line represents the error bound indicated by the RHS of Eq.~\eqref{eq:LeCam}. Upper and lower dotted lines denote the maximum and minimum error bounds, respectively. (b) Mean and standard deviation of absolute changes in $N$. }
    \label{fig:select_tau}
\end{figure}

\clearpage

  \begin{figure}[b]
    \centering
       \includegraphics[width=.9\columnwidth]{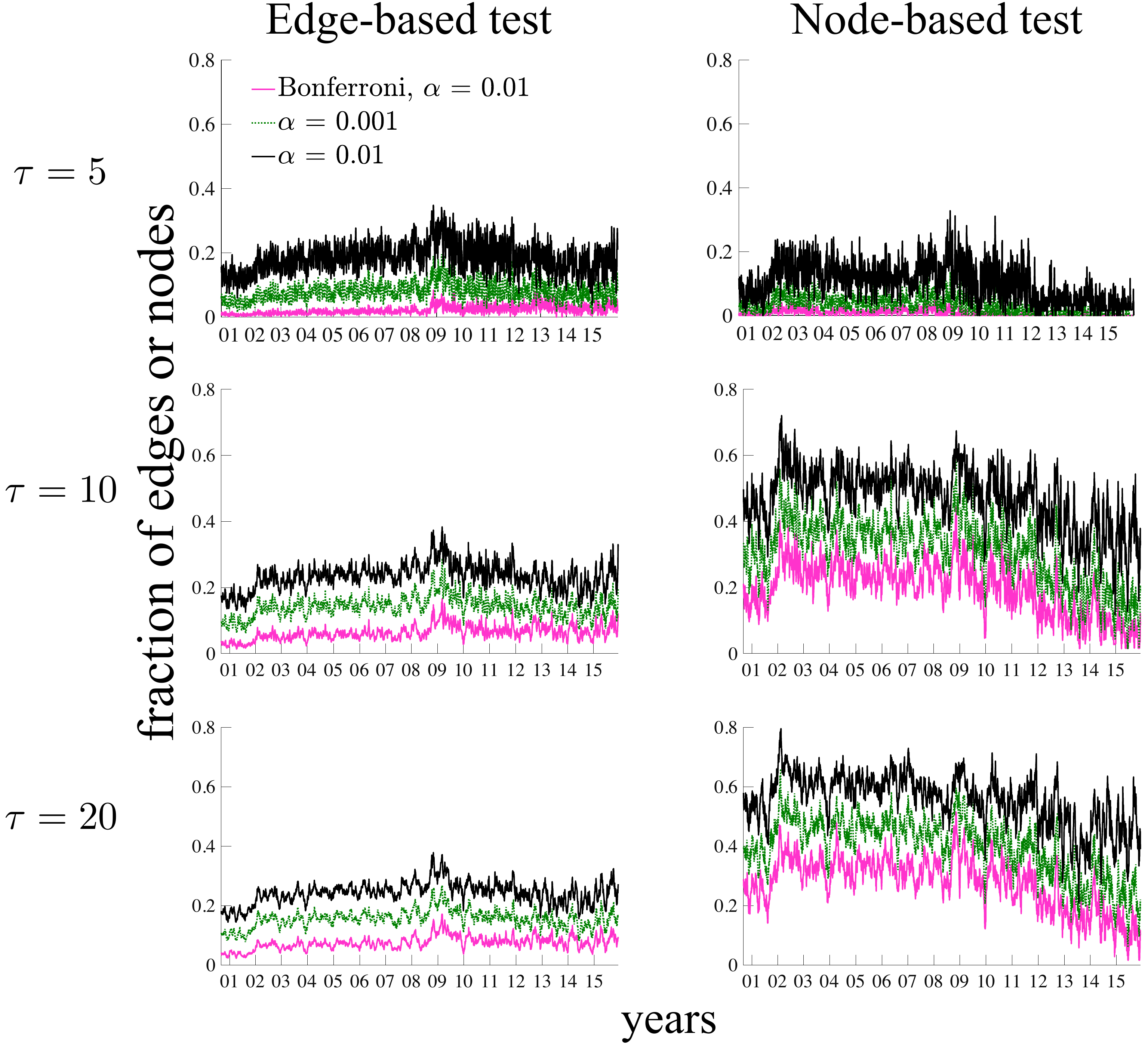}
           \caption{\new{Effects of varying aggregate length $\tau$. Left and right columns present the fraction of significant ties and the fraction of relationship-dependent banks, respectively. A rolling window is used.}}
    \label{fig:frac_pref_robustcheck}
\end{figure}

\begin{figure}[b]
    \centering
       \includegraphics[width=.7\columnwidth]{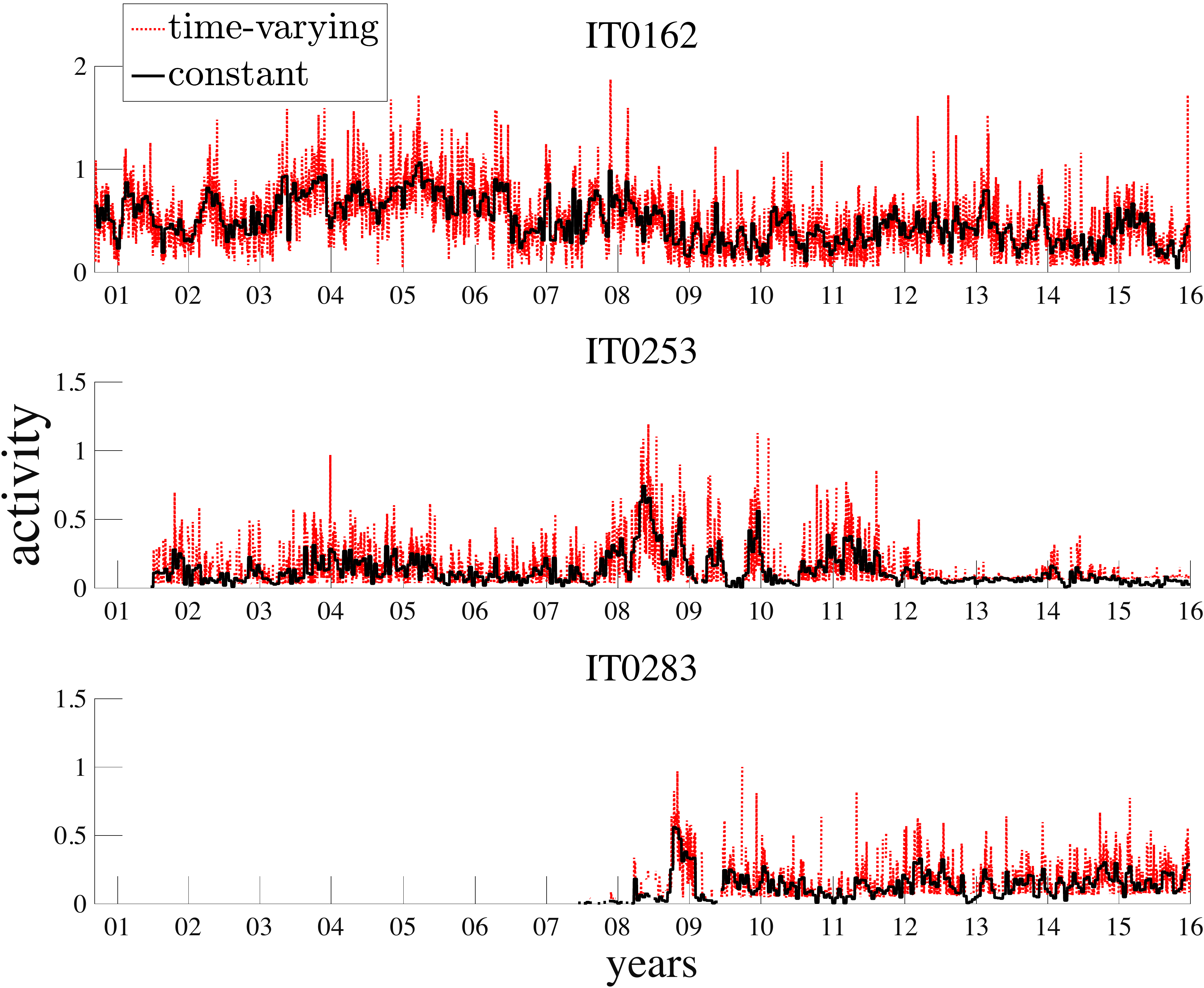}%
           \caption{\new{Comparison between constant and time-varying activity levels of a bank. Bank ID is annotated at the top of each panel. The three banks are ranked 1st (top), 50th (middle) and 100th (bottom) among 308 banks in terms of the number of participating days.} }
    \label{fig:var_const_activity}
\end{figure}

\end{document}